\documentclass[9pt,twocolumn,twoside]{pnas-new_arxiv}
\templatetype{pnasresearcharticle_arxiv} % Choose template 
\usepackage{url}
\usepackage{graphicx} 
\usepackage{widetext}%%%NOTE %%%% this is a modified (to remove lines) version by TK Mar31, 2017%%%%%

\newcommand{\Np}{N_{\rm P}}
\newcommand{\Npml}{N_{\rm P,ML}}

\title{Social dynamics of financial networks}

\author[a]{Teruyoshi Kobayashi}
\author[b,c,d,1]{Taro Takaguchi} 

\affil[a]{Graduate School of Economics, Kobe University, Kobe, Japan}
\affil[b]{National Institute of Information and Communications Technology, Tokyo, Japan}
\affil[c]{National Institute of Informatics, Tokyo, Japan}
\affil[d]{JST, ERATO, Kawarabayashi Large Graph Project, Tokyo, Japan}
\correspondingauthor{\textsuperscript{1}Corresponding author. E-mail: taro.takaguchi@nict.go.jp}

\begin{abstract}
\textbf{Abstract}\\
The global financial crisis in 2007--2009 demonstrated that systemic risk can spread all over the world through a complex web of financial linkages, yet we still lack fundamental knowledge about the evolution of the financial web.
 In particular, interbank credit networks shape the core of the financial system, in which a time-varying interconnected risk emerges from a massive number of temporal transactions between banks. The current lack of understanding of the mechanics of interbank networks makes it difficult to evaluate and control systemic risk.
Here, we uncover fundamental dynamics of interbank networks by seeking the patterns of daily transactions between individual banks.
We find stable interaction patterns between banks from which distinctive network-scale dynamics emerge.
In fact, the dynamical patterns discovered at the local and network scales share common characteristics with social communication patterns of humans.
To explain the origin of ``social'' dynamics in interbank networks, we provide a simple model that allows us to generate a sequence of synthetic daily networks characterized by the observed dynamical properties. The discovery of dynamical principles at the daily resolution will enhance our ability to assess systemic risk and could contribute to the real-time management of financial stability. 
  \end{abstract}

\dates{\today}
\begin{document}

% Optional adjustment to line up main text (after abstract) of first page with line numbers, when using both lineno and twocolumn options.
% You should only change this length when you've finalised the article contents.
\verticaladjustment{-2pt}

\maketitle
\thispagestyle{firststyle}
\ifthenelse{\boolean{shortarticle}}{\ifthenelse{\boolean{singlecolumn}}{\abscontentformatted}{\abscontent}}{}

%systemic risk の重要性
\section*{Introduction}
Financial systemic risk is one of the most serious threats to the global economy. The global financial crisis of 2007--2009 showed that a failure of one bank can lead to a financial contagion through a complex web of financial linkages, which are created by everyday transactions among financial institutions~\cite{Brunnermeier2009JEP,Allen2010}. Even after the crisis, many countries have experienced a prolonged recession, the so-called Great Recession, showing that the social cost of a financial crisis can be enormous~\cite{Mishkin2011NBER,Atkinson2013cost}.
Evaluating and controlling systemic risk has therefore been recognized as one of the greatest challenges for interdisciplinary researchers across different fields of science~\cite{May2008Nature,Schweitzer2009Science,Helbing2013nature,Battiston2016Science}.   

%interbank の重要性
In the modern financial system, interbank markets play a fundamental role, in which banks lend to and borrow from each other (hereafter, we refer to all types of financial institutions as ``banks" for brevity).
 Lending and borrowing in interbank markets are necessary daily tasks for banks to smoothen their liquidity management~\cite{Beaupain2008}, but at the same time, they also form the center of a global web of interconnected risk;  shocks to interbank markets may spill over to other parts of the global financial system through the financial linkages to which banks are connected~\cite{Cifuentes2005JEEA,Schweitzer2009Science,Huang2013SciRep}. 
%dynamic analysis の重要性
 Many previous studies attempt to assess systemic risk by simulating different scenarios of cascading bank failures on both real~\cite{Upper2004EER,Elsinger2006,Lelyveld2006IJCB,Cont2013,Bardoscia2017Natcom} and synthetic interbank credit networks~\cite{Nier2007,GaiKapadia2010,Haldane2011Nature,Huang2013SciRep,Tedeschi2013plosone,Brummitt2015PRE,Burkholz2016physicaD}. Studies of financial cascades based on synthetic networks often assume a particular static structure, such as random~\cite{GaiKapadia2010,Haldane2011Nature}, bipartite~\cite{Huang2013SciRep,Caccioli2015}, and multiplex structures~\cite{Brummitt2015PRE,Burkholz2016physicaD}, 
 successfully revealing that the structural property affects the likelihood of default cascades to a large extent.
However, since the great majority of real-world interbank transactions are in fact overnight~\cite{Beaupain2008,Iori2015}, interbank networks should be treated as dynamical systems with their structure changing on a daily basis. This temporal nature of real interbank networks inevitably limits the practical usefulness of the conventional static approach to systemic risk. Nevertheless, we still have little knowledge about how the structural characteristics of daily networks evolve over time. It has long been believed that the dynamics of interbank networks is random and thus has no meaningful regularity at the daily scale~\cite{Finger2013}.  How the current network structure is correlated with the past structures, or more specifically, how banks choose current trading partners based on their trading history, is still unknown. The ambiguity of the structural dynamics of interbank networks itself may also become a source of systemic risk by veiling the complexity of interconnectivity~\cite{Battiston2016PNAS}. The current lack of studies on the mechanics of real interbank networks is in stark contrast to the abundance of research on their static property~\cite{Boss2004,Upper2004EER,Soramaki2007physicaA,Iori2008JEDC,Imakubo2010BOJ,Cont2013}.

  %結果の概要、社会的意義
  The main aim of this work is to uncover fundamental dynamics governing real interbank networks at both local and system-wide scales. For this purpose, we first seek dynamical regularities that would characterize interaction patterns between individual banks by looking at millions of overnight transactions conducted in the Italian interbank market during 2000--2015~\cite{emidHP}. We discover that there exist explicit interaction patterns that rule daily bank-to-bank transactions, which turn out to be essentially the same as the patterns characterizing human social communication; that is, banks trade with their partners in the same way that people interact with friends through phone calls and face-to-face conversations~\cite{Cattuto2010PLOS,Starnini2013PRL}. In fact, those ``social'' interaction patterns of banks have been surprisingly stable over time, even amid the global financial crisis. On top of local interactions between banks, there emerges a system-wide scaling relationship between the numbers of banks and transactions, just as the number of phone-call pairs scales superlinearly with the size of population~\cite{Schlapfer2014}.

 To explain the origin of social dynamics in interbank networks, we develop a model that generates a sequence of synthetic daily networks from which all the observed dynamical patterns simultaneously emerge at both local and network scales.
Our discovery of the fundamental mechanism underpinning the daily evolution of interbank networks will enhance the predictability of systemic risk and provide an important step toward the real-time management of financial stability.

\section*{Results}

The dataset to be analyzed in this work is the time series of daily networks identified from the time-stamped data of interbank transactions conducted in the Italian interbank market during 2000--2015 (Materials and Methods: Data). 
The daily interbank networks have directed edges originating from lending banks to borrowing banks. One may regard the amount of funds transferred from a lender to a borrower as edge weights, but here we regard the daily networks as unweighted, since we found that the dynamics of edge weights can be decoupled from the edge dynamics themselves (see Supplementary Materials (SM) for the analysis of edge weights).

\begin{figure}[t]
\begin{center}
           \includegraphics[width=.9\columnwidth]{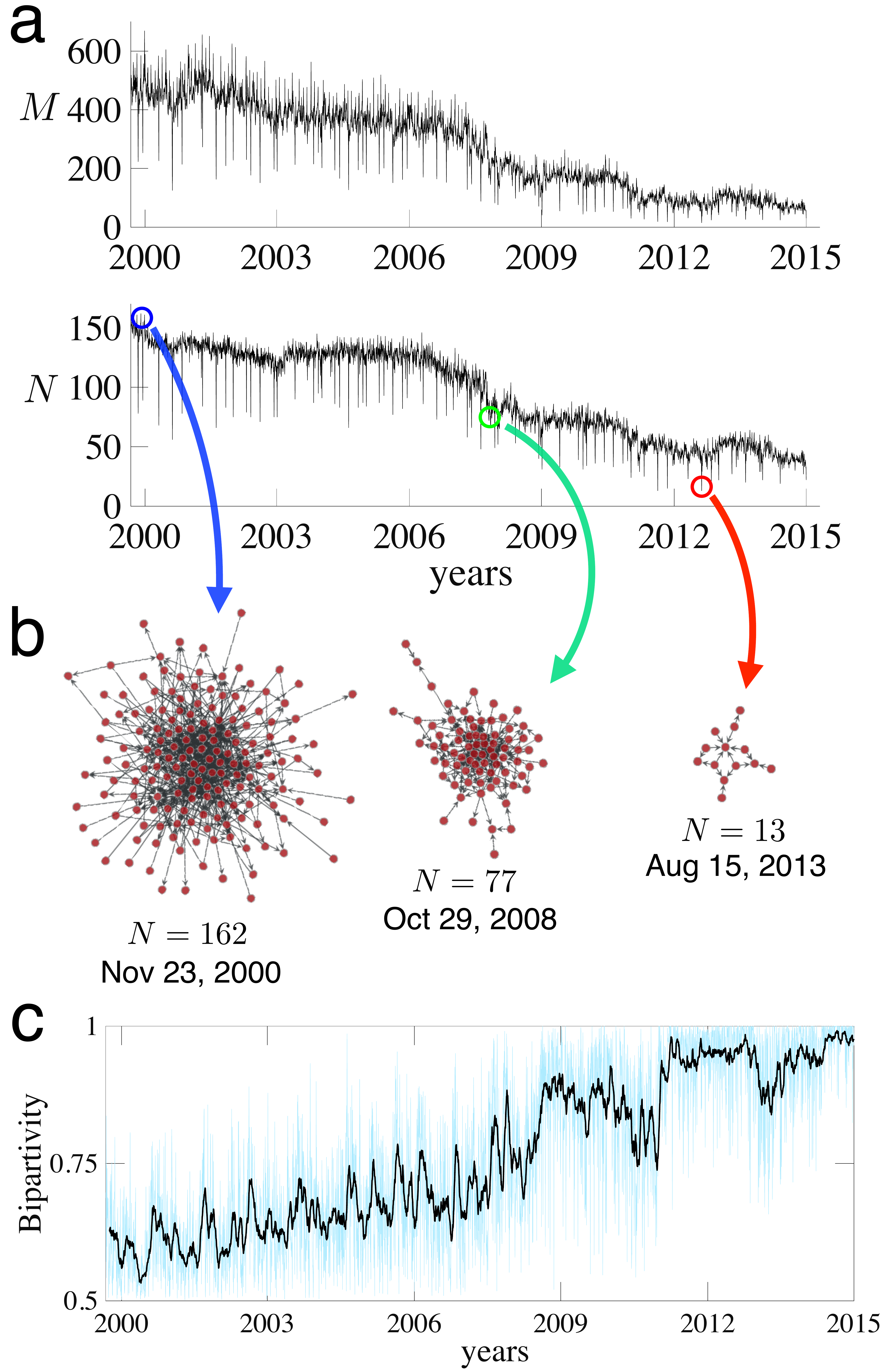}
    \end{center}
    \caption{Time series of daily networks. ({\textbf a}) Daily time series of the number of edges $M$ (upper) and the number of banks $N$ (lower). Most of the downward spikes in $N$ and $M$ are due to the national holidays in Italy. ({\textbf b}) Visualization of the largest (left), a middle-sized (middle) and the smallest (right) daily networks (visualized by graph-tool~\cite{graph-toolHP}) ({\textbf c}) Time series of bipartivity. Bipartivity is a measure of bipartite structure that takes $1$ if the network is fully bipartite and $0.5$ if a complete graph~\cite{Estrada2005PRE}. Black line represents the moving average with 20-day smoothing window.}
 \label{fig:netvisualize_data}
\end{figure}

An important general observation regarding the dynamics of daily networks is that both the network size $N$ and number of edges $M$  have followed downward trends (Fig.~\ref{fig:netvisualize_data} a and b). This led the networks closer to a bipartite structure between \textit{pure lenders} and \textit{pure borrowers}~\cite{Barucca2015ecb} (Fig.~\ref{fig:netvisualize_data}c and Table~\ref{tab:frac_banktype}), entailing a rapid turnover in the set of banks participating in each daily network (see the turnover rate in Table~\ref{tab:summary_stats} and Fig.~\ref{fig:turnover}). Table~\ref{tab:summary_stats} summarizes the basic statistics, in which we divide the entire sample period into three subsample periods to observe whether a structural change around the global financial crisis in 2007--2009 is present.

\begin{table}[b]
            \caption{Summary statistics of the daily interbank networks}
    \begin{tabular}{lcccc}
    \hline
    \hline
       & All & \hspace{.1cm}2000--2006 & \hspace{.1cm}2007--2009 & \hspace{.1cm}2010--2015 \\ 
       \hline
        \# days &3,922  &1,618  &767& 1,537  \\
       $\overline{N}$  &94.23 &129.69  &98.92 & 54.56 \\
        $\overline{M}$ &302.76  &466.35 &304.89 & 129.48  \\
        Turnover rate  &0.22  & 0.18 &0.22& 0.26  \\ 
         Bipartivity & 0.77 & 0.64 & 0.78 & 0.92 \\
        \hline
    \end{tabular}\\
 \footnotesize{$\overline{N}$ and $\overline{M}$ denote the average numbers of active banks and edges in the daily networks, respectively.  The turnover rate is the average of the Jaccard distance $1-|I_{t}\cup I_{t-1}|/|I_{t}\cap I_{t-1}|$, where $I_{t}$ is the set of active banks on day $t$. See caption of Fig.~\ref{fig:netvisualize_data}c for a description of bipartivity.}
    \label{tab:summary_stats}
\end{table}

\subsection*{Dynamical patterns of daily networks}

\begin{figure}[t]
\begin{center}
         \includegraphics[width=.95\columnwidth]{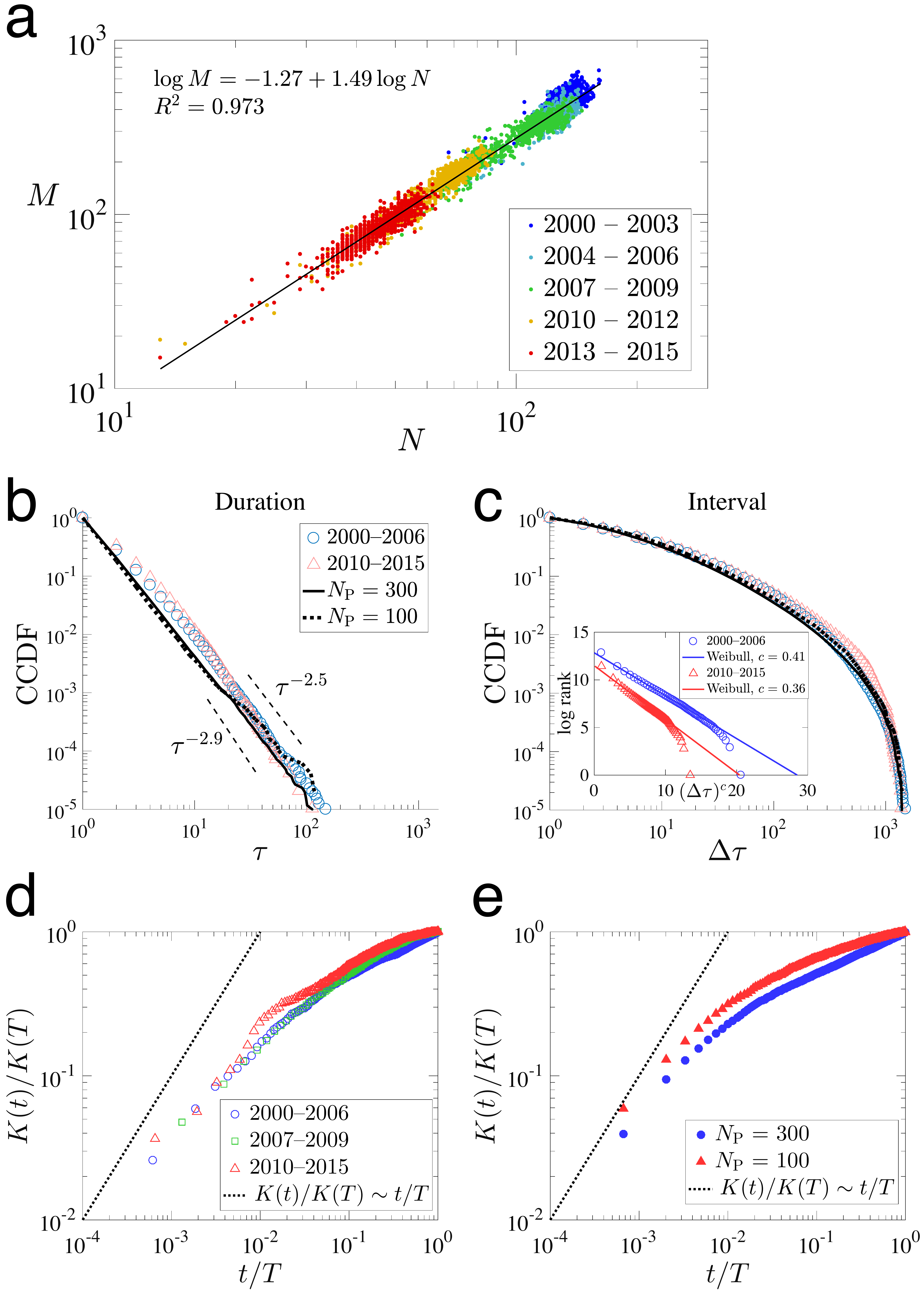}
    \end{center}
    \caption{Dynamical patterns of daily networks. ({\textbf a}) $M$ is a power-law function of $N$. A dot corresponds to a daily network, and solid line represents a log-linear regression estimated by the ordinary least squares. ({\textbf b}) Distribution of consecutive trading days, $\tau$, for a bank pair, aggregated over all pairs. Blue circles and red triangles indicate the empirical distributions for 2000--2006 and 2010--2015, respectively. Solid and dotted lines represent simulated distributions with $\Np = 300$ and $100$, respectively.  ({\textbf c}) Distribution of interval time, $\Delta \tau$, between two consecutive transactions for a bank pair, aggregated over all pairs. \textit{Inset}: log-rank plot of $(\Delta\tau)^c$ indicated by a Weibull distribution with estimated parameter $c$ (symbols), and its theoretical values (lines). ({\textbf d}) Time series of the empirical aggregate degree $K(t)$ (averaged over all banks) normalized by its terminal value $K(T)$ and ({\textbf e}) the simulated aggregate degree.}
 \label{fig:durint_scaling_model}
\end{figure}

The downward trends in $N$ and $M$, along with the intermittent spikes, left a broad range of daily combinations $(N,M)$, which allows us to ask how the number of financial linkages is dynamically constrained by the number of banks.
In fact, there arises a clear superlinearity, $M\propto N^{1.5}$ (Fig.~\ref{fig:durint_scaling_model}a). This suggests that the average degree of a daily network increases with order $\sqrt{N}$, or $\langle k\rangle \propto \sqrt{N}$.
It should be noted that the fact that $M$ is given as a power-law function of $N$ is similar to a widely observed phenomenon in social networks, called superlinear scaling, in which the number of edges scales superlinearly with the number of nodes across different locations~\cite{Bettencourt2007,Bettencourt2013,Pan2013,Schlapfer2014}.

In addition to the macroscopic dynamics of $N$ and $M$, we also find characteristic properties of the microdynamics of individual edges: edge duration and interval time.
We define duration $\tau$ as the number of successive business days on each of which a bank pair performs at least one transaction.
Aggregated over all trading pairs, $\tau$ follows a power-law distribution whose complementary cumulative distribution function (CCDF) has an exponent between $2.5$ and $2.9$ (Fig.~\ref{fig:durint_scaling_model}b and Fig.~\ref{fig:durint_data} in SM. The exponents are estimated using the method proposed in~Refs.~\cite{Clauset2009Siam,ClausetHP}).
Similar power-law distributions are observed when we redefine $\tau$ as the duration of individual banks' successive trading activity either for lending, borrowing, or both  (Fig.~\ref{fig:durint_model_node_link} in SM).

On the other hand, the interval time $\Delta \tau$ for a bank pair is defined as the interval length between two consecutive transactions during which the bank pair performs no transactions.
In contrast to $\tau$, $\Delta \tau$ does not follow a power-law distribution, while it still shows a long-tailed behavior (Fig.~\ref{fig:durint_scaling_model}c).
The interval distribution fits well with a Weibull distribution up to a certain cut-off level (Fig.~\ref{fig:durint_scaling_model}c, \textit{Inset}. See section \ref{sec:weibull_fit} of SM for details on the fitting method~\cite{Sornette2006Book,Weibull1951}). 

We observe that the distributions of $\tau$ and $\Delta\tau$ have been quite stable throughout the whole data period. This observation is notable, not only because $N$ and $M$ continually fluctuate at a daily resolution over the course of the decreasing trends (Fig.~\ref{fig:netvisualize_data}a), but also because a large fraction of the set of participating banks changes day to day (Fig.~\ref{fig:turnover}).
The high metabolism of the interbank market suggests that the stationarity of $\tau$ and $\Delta\tau$ is not necessarily attributed to the presence of steady relationships between particular banks.

 While the dynamics of individual edges in daily networks is shown to follow particular patterns, it would also be meaningful to see the dynamics of aggregated edges (i.e., edges of aggregated networks).  
We focus on the aggregated degree $K(t)$, defined by the average cumulative number of unique trading partners up to time $t$~\cite{Song2010NatPhys,Starnini2013PRL}.
The normalized aggregate degree, defined by $K(t)/K(T)$, grows sublinearly in time (Fig.~\ref{fig:durint_scaling_model}d), meaning that the rate at which banks find a new partner tends to decrease over time. Note that such sublinear growth patterns are also reported for the mobility pattern of mobile-phone users~\cite{Song2010NatPhys} and for contact networks of human individuals formed via face-to-face interactions~\cite{Starnini2013PRL}.

\subsection*{Model of daily network evolution}\label{sec:model}

The above findings show that the dynamical patterns of interbank transactions are robust across different periods, which leads us to consider that a universal mechanism generating daily interbank networks might exist.
Here, we show that the emergence of these regularities can be reconstructed by a dynamical generalization of the fitness model~\cite{Caldarelli2002PRL,DeMasi2006PRE} (see Materials and Methods: Model).

First, we show that variations in the system size of a simple fitness model can explain the empirical superlinear relation $M \propto N^{1.5}$. For ease of exposition, suppose for the moment that the networks are undirected.
In the fitness model, fitness value $a_i \in [0,1]$ is assigned to bank~$i$ $(1 \leq i \leq \Np)$, where $\Np$ represents the potential market size, given by the number of banks that may perform transactions during a day.
In the context of interbank transactions, fitness value $a_i$ can be interpreted as the activity level, or willingness, of bank $i$ to trade.
The probability that an edge is formed between $i$ and $j$ is given by $p_{ij} = (a_i a_j)^\alpha$ $(\alpha \geq 1)$. 
For each network generated by this rule, $N$ and $M$ denote the number of active banks with at least one edge (thus $N \leq \Np$) and the total number of edges, respectively.

 \begin{figure}
\begin{center}
      \includegraphics[width=.85\columnwidth]{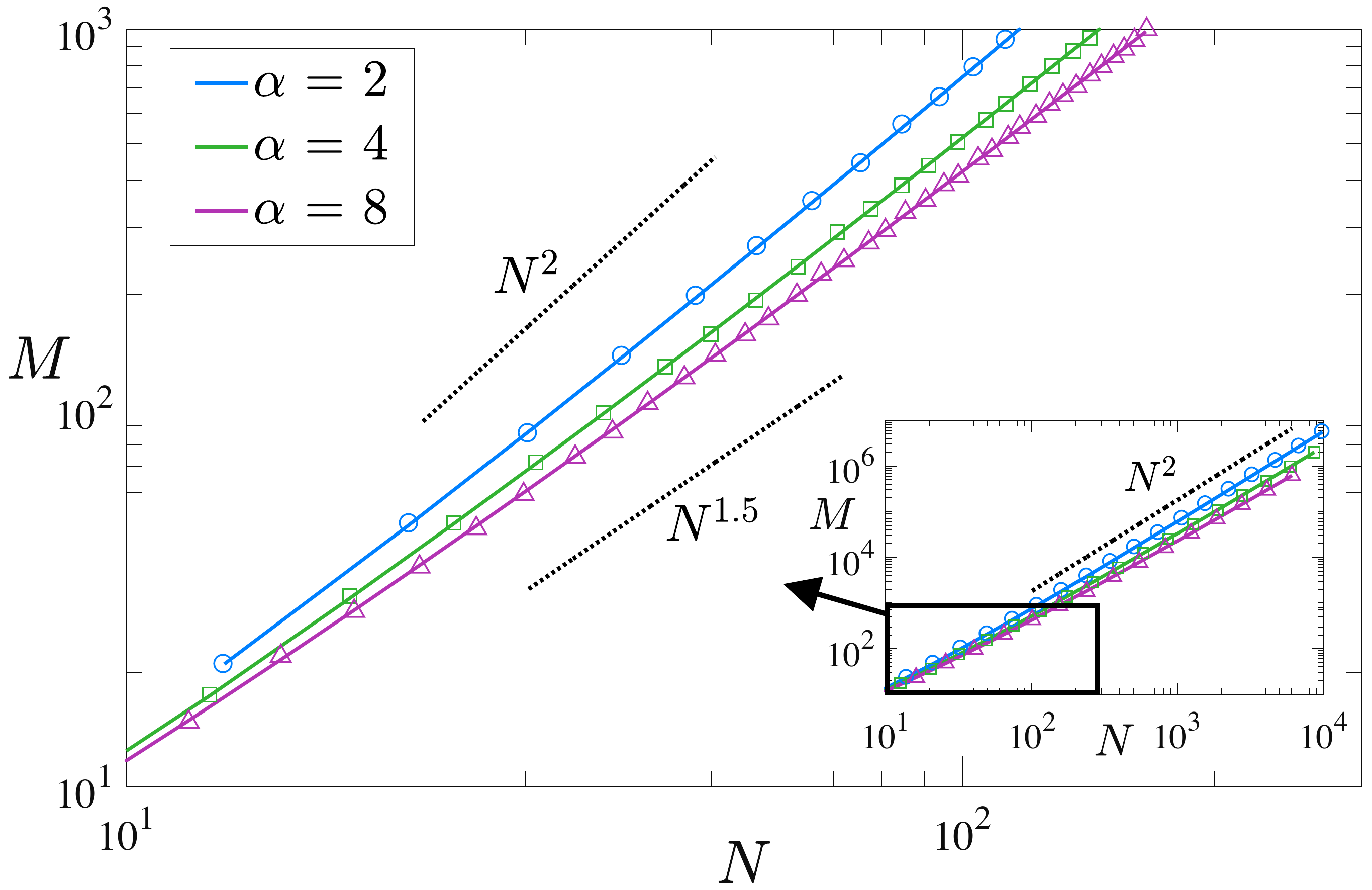} 
    \end{center}
    \caption{Superlinear relation in the fitness model. Solid lines denote the theoretical values indicated by \eqref{eq:NM}, while symbols represent the corresponding simulation results (average of 500 runs). Scaling relation $M\propto N^\beta$ with $1 < \beta < 2$ emerges by varying $\Np$ when $\Np$ is small (approximately $20 \leq \Np \leq 300$). \textit{Inset}: increasing $\Np$ to 10000 restores quadratic scaling $M\propto N^2$ since $q_0\approx 0$ and thereby the finite-size effect disappears.}
    \label{fig:NM_theory}
\end{figure}

By generating model networks with $\Np$ varying from 20 to 300 for a given $\alpha\in [2,8]$, there arises a scaling relation $M \propto N^\beta$ with $1 < \beta < 2$  (symbols in Fig.~\ref{fig:NM_theory}).
In previous studies~\cite{Caldarelli2002PRL,Boguna2003PRE,DeMasi2006PRE}, a theoretical analysis of the fitness model predicted $M \propto N^2$, which differs from both our empirical observations (Fig.~\ref{fig:durint_scaling_model}a) and numerical simulations (Fig.~\ref{fig:NM_theory}).
In fact, this discrepancy is explained by the presence of isolated banks. In this model, the probability of a bank being isolated (i.e., no edges attached), defined by $q_0$, is given by a function of $\Np$:
\begin{align}
q_0(\Np)= \frac{1}{\alpha} (\alpha + 1)^{\frac{1}{\alpha}} \left[ {\rm B}\left( \Np, \frac{1}{\alpha}\right) - {\rm B}_{1-\frac{1}{\alpha+1}} \left( \Np, \frac{1}{\alpha}\right)\right] ,
\label{eq:q0}
\end{align}
where ${\rm B}(x,y)$ and ${\rm B}_z(x,y)$ are beta and incomplete beta functions, respectively (see section~\ref{sec:analytical} for derivation).
Consequently, $N$ and $M$ are given by
\begin{align}
\begin{cases}
N = (1-q_0(\Np))\Np,\\
M = \langle (a_i a_j)^\alpha \rangle \frac{\Np(\Np-1)}{2}.
\end{cases}
\label{eq:NM}
\end{align}
Since $q_0(\Np) \to 0$ as $\Np \to \infty$, $N \simeq \Np$ and 
$M \simeq  \langle (a_i a_j)^\alpha \rangle N(N-1)/2 \propto N^2$ hold true for a sufficiently large $\Np$, which recovers the quadratic scaling shown in the previous studies~\cite{Caldarelli2002PRL,Boguna2003PRE,DeMasi2006PRE}.
However, for the range of network sizes observed from the data, $q_0(\Np)$ is not negligible. A combination of $(N,M)$ derived from \eqref{eq:NM} for given values of $\Np$ perfectly fits the simulation result (lines in Fig.~\ref{fig:NM_theory}) .

While the superlinear relation $M \propto N^{1.5}$ can be explained by variations in $\Np$, this simple model cannot reproduce the distributions of $\tau$ and $\Delta \tau$ (Fig.~\ref{fig:durint_scaling_model} b and c) and the sublinear growth of $K(t)$ (Fig.~\ref{fig:durint_scaling_model}d) since these characteristics come from the effect of memory in the formation of links between banks.
To capture the memory effect, we introduce a fluctuation in fitness $a_i$. 
We assume that at the beginning of day $t+1$, $a_{i,t+1}$ is updated according to a random walk process or reset to a random value between 0 and 1 with probability $h(a_{i,t})$ (see Materials and Methods: Model). 
The reset probability is intended to capture the metabolism of interbank markets, in which some banks exit the market after continuous transactions, whereas other banks enter after a long resting periods (e.g., due to a change in the strategy of liquidity management).

    \begin{figure}
\begin{center}
      \includegraphics[width=.95\columnwidth]{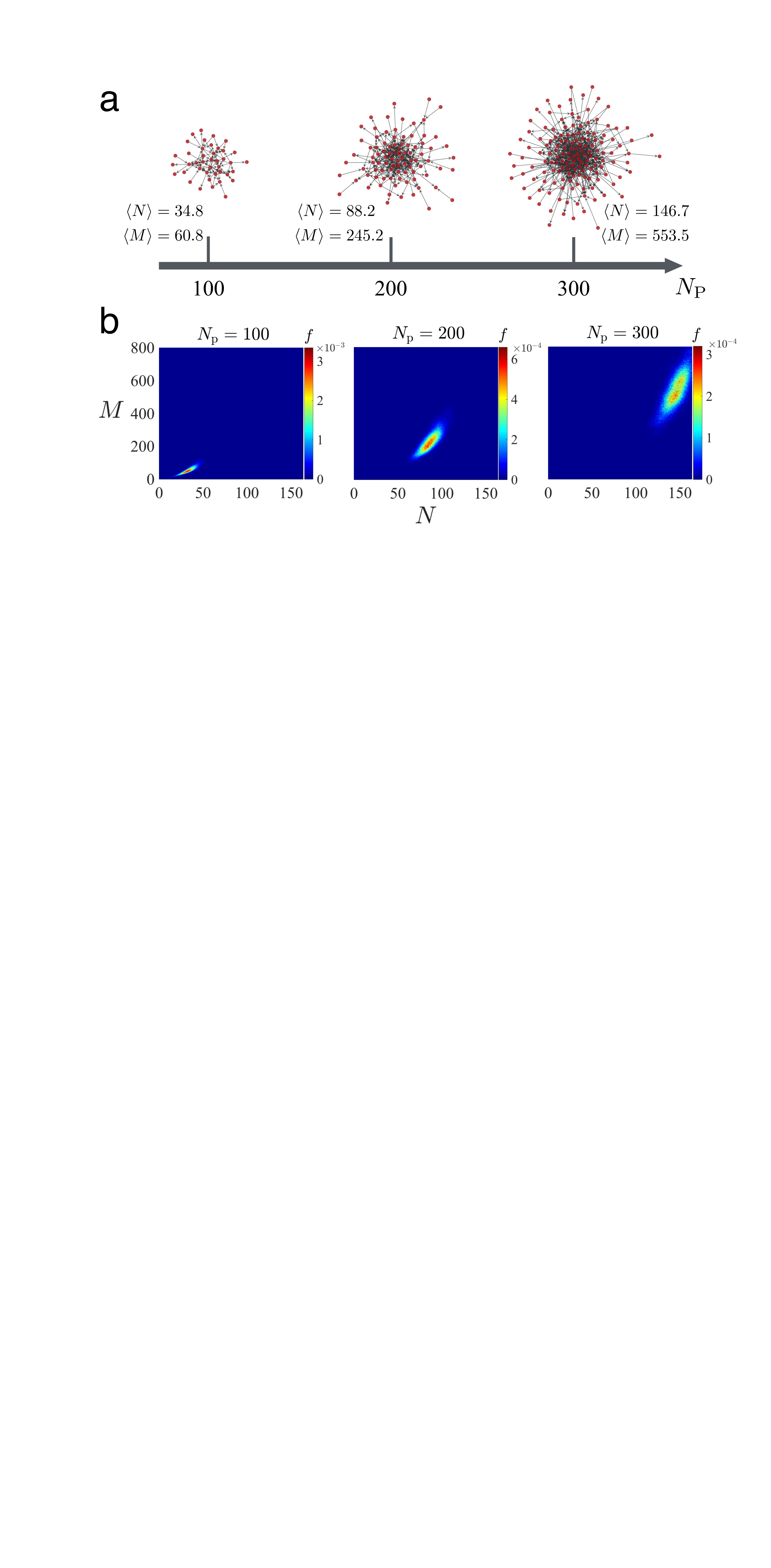} 
    \end{center}
    \caption{Generation of model networks. ({\textbf a}) Model networks with $\Np =100, 200,$ and $300$ (visualized by graph-tool~\cite{graph-toolHP}). The average values $\langle N\rangle$ and $\langle M\rangle$  for each $\Np$ are also shown. ({\textbf b}) Joint conditional probability function $f(N, M | \Np)$ (color bar) for a given $\Np$.}
\label{fig:netvisual_model}
\end{figure}
  
We find that the simulated distributions of duration $\tau$ and interval $\Delta \tau$ for pairwise transactions replicate the empirical distributions for a given $\Np$ (see lines in Fig.~\ref{fig:durint_scaling_model} b and c).
We confirmed that the model can robustly reproduce the duration and interval distributions under different parameter settings (Figs.~\ref{fig:scaling_diffalpha} b and c, \ref{fig:duration_diff_resetprob} and \ref{fig:interval_diff_resetprob}). In addition, the growth pattern of the normalized aggregate degree $K(t)/K(T)$ is successfully reproduced (Fig.~\ref{fig:durint_scaling_model}e). We also evaluate the model fit for other dynamical properties such as the degree distribution and the strength as a function of degree~\cite{Gautreau2009PNAS} (Figs.~\ref{fig:degreedist}--\ref{fig:strength_model}).

It should be noted that while the activity level $a_{i,t}$ fluctuates with time independently from other banks' activity levels, the model ensures that the size and the structure of generated networks are stationary for a given $\Np$. In reality, however, the evolution of daily networks show a decreasing trend (Fig.~\ref{fig:netvisualize_data} a and b) and the network size varies from day to day owing to various external factors (e.g., shifts in monetary policy~\cite{Barucca2015ecb} and the seasonality of money demand due to the national holidays and/or the reserve requirement system~\cite{Beaupain2008}). In the model, the averages of the network size $\langle N\rangle$ and the number of edges $\langle M\rangle$ are controlled by tuning parameter $\Np$ (Fig.~\ref{fig:netvisual_model}a).

 \subsection*{Fitting model to the data}

\begin{figure}[t]
\begin{center}
         \includegraphics[width=.95\columnwidth]{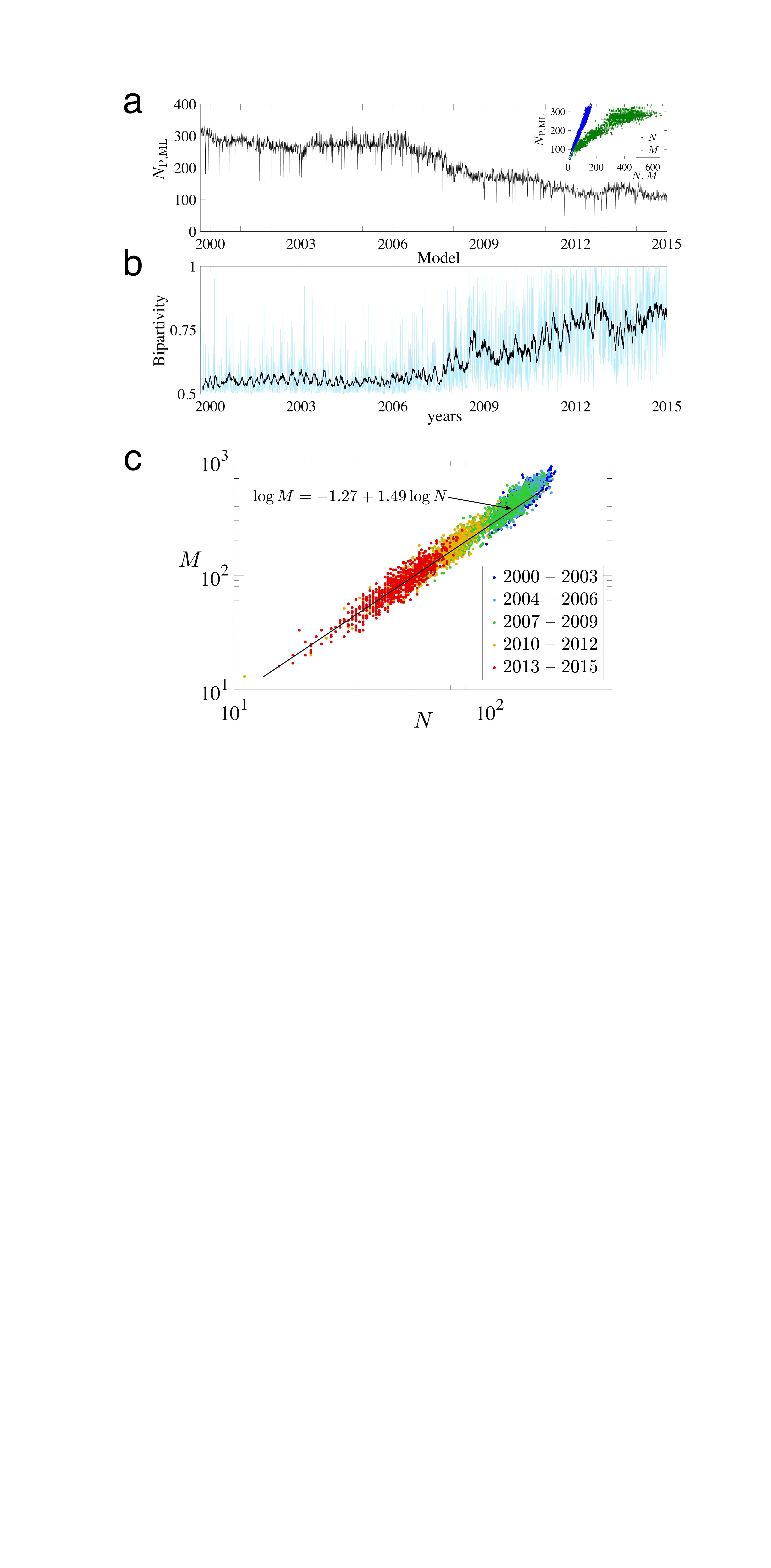} %scaling_model_edge.fig
    \end{center}
    \caption{Dynamic characteristics of fitted daily networks. ({\textbf a}) Time series of $\Npml$. \textit{Inset}: Scatter plot of $\Npml$ against empirical $N$ and $M$. ({\textbf b}) Bipartivity of fitted daily networks. Black line represents the moving average with 20-day smoothing window. ({\textbf c}) Scatter plot of $N$ and $M$ of fitted daily networks. A dot corresponds to a day. Solid line illustrates the empirical regression line identical to that shown in Fig.~\ref{fig:durint_scaling_model}a.}
\label{fig:estimated_Np}
\end{figure}

In practice, the influence of exogenous factors that would affect the network size might vary daily. Since in the model the average network size is controlled by tuning parameter $\Np$, we need to estimate the daily sequence of $\Np$ to reconstruct time series of empirical daily networks. Here, we take the following steps.
First, we generate sufficiently many instances of synthetic networks for a given $\Np$ to compute the histogram of $(N, M)$ (Fig.~\ref{fig:netvisual_model}b).
Generating networks over a sufficiently broad range of $\Np$ provides a conditional probability function $f(N, M| \Np )$ that would cover the range of $N$ and $M$ observed in the empirical networks (Fig.~\ref{fig:MLfunc_scaling}). 
Second, for a given empirical daily network with $(N^\prime, M^\prime)$, we choose a daily estimate of $\Np$, denoted by $\Npml$, such that $\Npml = {\rm argmax}_{\Np}f(N^\prime, M^\prime | \Np )$. 

 We find that simulated instances of $(N, M)$ concentrate tightly along the regression line of $M\propto N^{1.5}$ when we vary $\Np$ from $50$ to $350$ (Fig.~\ref{fig:MLfunc_scaling}).
This agreement holds true even under alternative parameter values within a reasonable range of variation (Figs.~\ref{fig:scaling_diffalpha}a and \ref{fig:scaling_diff_resetprob}).
The sequence of daily estimates of $\Npml$, each of which is based on the empirical combination $(N^\prime, M^\prime)$ of a day, exhibit a long-term downward trend consistent with the empirical sequence of $N$ and $M$ (Fig.~\ref{fig:estimated_Np}a).
Specifically, $\Npml$ is proportional to $N$ and increases nonlinearly in $M$ with saturation at $\Npml \approx 300$ (Fig.~\ref{fig:estimated_Np}a, \textit{Inset}).

A time series of model networks based on the daily estimates of $\Npml$ reproduces the tendency toward a perfect bipartite structure (Fig.~\ref{fig:estimated_Np}b), although we have not explicitly modeled how the network structure should change with $\Np$. The tendency toward a bipartite structure may reflect the fact that the average degree of generated networks becomes smaller as the size of network shrinks.  
Indeed, the relationship between $N$ and $M$ across the fitted daily networks explains the emergence of superlinearity (Fig.~\ref{fig:estimated_Np}c), which indicates that $M \propto N^{1.5}$ or $\langle k\rangle \propto \sqrt{N}$.%\tk{直前でaverage degree の減少について言っているので、平均も併記しました。}

\section*{Discussion}
The time series of daily networks reveal many dynamical regularities encoded in millions of financial transactions.
An important finding is that the transaction patterns between banks are similar to the social communication patters of humans, which have been observed at higher temporal resolutions (typically 20$\sim$60 seconds) than a daily resolution. For instance, a power-law scaling in the distribution of the interaction duration has been found in human contact networks, such as face-to-face conversation networks of individuals~\cite{Cattuto2010PLOS,Starnini2013PRL}. The sublinear growth pattern of aggregated degree has also been reported in the mobility pattern of mobile-phone users~\cite{Song2010NatPhys}. In addition, superlinear scaling at the network level (called ``urban scaling"~\cite{Bettencourt2007}) emerges in various social contexts, such as the relationship between the number of mobile connections and the population size of cities~\cite{Schlapfer2014}. These similarities between financial transaction patterns of banks and social communication patterns of humans strongly suggest that banks choose trading partners in the same manner as individuals decide whom to talk with. 
The discovered dynamical patterns are quite robust and hold true even amid the global financial crisis, suggesting that there is a universal mechanism connecting financial and social dynamics.

 The contribution of our work is not limited to the findings on the transition patterns of interbank networks.  
The model we propose here can be used as a generator of synthetic networks for studies of financial systemic risk. 
As is often the case, inaccessibility to empirical data on financial transactions forces academic researchers to use synthetic networks with limited empirical properties~\cite{Huang2013SciRep,Brummitt2015PRE} or to infer real network structure based on available partial information~\cite{Lelyveld2006IJCB,Squartini2013SciRep,Mastrandrea2014NJP}.
Our model provides a way to easily generate synthetic time series of networks that exhibit dynamical properties characterizing the daily evolution of real interbank networks. We hope that the current work will not only deepen our knowledge about the dynamic nature of interbank networks, but also help to improve the conventional approach of systemic-risk studies toward a more dynamic analysis.

 We leave three remaining issues that need to be addressed in future research. First, while the observed dynamical patterns are quite robust and seem universal given the similarity to social network dynamics, it is worth investigating whether those findings hold true in other countries as well.  Second, we might be able to find other dynamical patterns at different time resolutions such as intraday, weekly, and monthly. If that is the case, we need to see if those dynamical patterns found in different time scales are consistently explained by the current model. Finally, our finding reveals an independence of local interaction patterns of banks from a global-scale network evolution, such as the decreasing trend in network size. This implies that there is no feedback loop between micro and macroscopic phenomena, meaning that banks are not adaptive to their environments. Further research is needed to explain why such a decoupling phenomenon takes hold in financial networks.      

\section*{Materials and Methods}

\subsection*{Data}
The original time-stamped data are commercially available from e-MID SIM S.p.A based in Milan, Italy~\cite{emidHP}. The data contain all the unsecured euro-denominated transactions between financial institutions made via an online trading platform, e-MID. 
%As of 2006, the amount of transactions conducted via e-MID accounts for 17\% of all the unsecured Euro-denominated transactions~\cite{ECB2006}. 
We focus on overnight (labelled as ``ON") and overnight-large (``ONL") transactions. ON transactions refer to contracts that require borrowers to repay the full amount within one business day from the day the loans are executed. ONL transactions are a variant of the ON transactions, where the amount is no less than 100 million euros. 

The data processing procedure is as follows.
First, we extract transactions made between September 4, 2000 and December 31, 2015. The choice of the initial date is based on the introduction of the ONL category~\cite{Beaupain2008}.
This leaves us with 1,119,258 ON and 73,480 ONL transactions, which comprise $86\%$ of all the
transactions during that period.
Next, we transform all the ON and ONL transactions into a sequence of daily networks by applying the daily time window of 8:00--18:00~\cite{Iori2015}.
We then extract the transactions that belong to the largest weakly connected component of each daily network, which account for $99.3\%$ of all the daily transactions on average (the minimum is $78\%$). We referred to this component as daily network throughout the analysis. Multiple edges between two banks are simplified. In the end, we have 1,187,415 transactions conducted by 308 financial institutions over 3,922 business days.

\subsection*{Model}

The dynamic network formation proceeds by repeating the following two steps: (i) edge creation between banks and (ii) update of each bank's activity level $a_{i,t} \in [0,1]$, $1 \leq i \leq \Np$.
We consider three bank types: \textit{pure lenders}, \textit{pure borrowers}, and \textit{bidirectional traders}. Pure lenders (pure borrowers) are the banks that may lend to (borrow from) but never borrow from (lend to) other banks. To take into account the fact that the interbank structure is almost perfectly bipartite when the network size is small (Fig.~\ref{fig:netvisualize_data}c), we assume that bidirectional traders may lend only to pure borrowers and borrow only from pure lenders.
The fraction of each bank type is given as $(f_{\rm B}, f_{\rm L}, f_{\rm D})=(0.56, 0.34, 0.1)$ for pure borrowers, pure lenders, and bidirectional traders, respectively, based on the empirical average (Table~\ref{tab:frac_banktype}). The type assigned to each bank is fixed throughout the simulation.

At the beginning of the edge-creation stage in day $t$, the interbank system comprises $\Np$ isolated banks without any edges.
Bank $i$ lends to bank $j$ (and thus a directed edge from $i$ to $j$ is formed) in day $t$ with probability $p_{ij,t}$, given by

\begin{align}\label{eq:trading_prob}
    p_{ij,t} \equiv
    \begin{cases} 
        \left(a_{i,t}a_{j,t} \right)^{\alpha}, & \text{if  $i\notin B$, $j\notin L$, and $\{ i,j \}\not\subset D$,}\\
      0 & \text{otherwise},
     \end{cases}
    \end{align} 
where $L$, $B$, and $D$ denote the sets of pure lenders, pure borrowers, and bidirectional traders, respectively.
After applying this procedure to every combination of $(i,j)$, we remove all the edges and move on to day $t+1$. 

At the beginning of day $t+1$, the activity level of bank $i$ is updated as follows.
With probability $h(a_{i,t})$, $a_{i,t+1}$ is reset to a random value drawn from the uniform distribution on $[0,1]$.
With probability $1-h(a_{i,t})$, the activity level is updated according to a random walk process on the unit circle, given by
\begin{align}
a_{i,t+1} &= |\cos{\theta_{i,t+1}}|,  \label{eq:a} \\ 
\theta_{i,t+1} &= \theta_{i,t} + 2\pi \varepsilon_{i,t+1},\label{eq:theta}
\end{align} 
where $\theta_{i,t+1}$ is a random-walk variable that describes the angle on the unit circle (see Fig.~\ref{fig:schematic_circle} for a schematic). Since an activity level must be on $[0,1]$, $a_{i,t+1}$ is given by the absolute values of $\cos{\theta_{i,t+1}}$. $\varepsilon_{i,t+1}$ is a random variable uniformly distributed on $[-0.002,0.002]$. The initial value for the angle is set such that $\theta_{i,0} = \arccos(a_{i,0})$, where $a_{i,0}$ is drawn from the uniform distribution on $[0,1]$. The above two steps, edge creation and activity updating, are repeated until we reach the predefined terminal date $T$.

The introduction of stochastic variable $\varepsilon$ is meant to capture fluctuations in individual banks' daily liquidity demand, which can lead to a turnover of participating banks (Fig.~\ref{fig:turnover}). Without a variability of $\varepsilon$ (i.e., if $a$ is fixed), the metabolism of the model interbank market would  be unrealistically low. 
%\del{An advantage of using a circular random-walk model, as opposed to a one-dimensional random walk with reflective boundary conditions, is that the distribution of $a$ will be skewed to large values, whereas $a$ would be uniformly distributed under a one-dimensional random walk. 
%This is important, because the stickiness of $a$ to a high-activity region contributes to reproducing the long-tailed duration distribution of pairwise transactions by allowing two high-activity banks to interact for a prolonged time. A long-lasting relationship will nevertheless end at some point in time, since there is a reset probability $h(a)$.}
The reset probability function is specified as  $h(a_i) \equiv c_{1}^{-1}a_{i}^{c_{2}}$.

In total, the model has four parameters: $\Np$, $\alpha$, $c_1$, and $c_2$.
Parameter $\Np$ is a key parameter of the model and we explain its role in the main text.
The other parameters, $\alpha$, $c_1$, and $c_2$, affect the structure of networks through the edge-creation probability \eqref{eq:trading_prob}. 
 We find that the combination $(\alpha, c_1, c_2)=(4, 2000,2)$ gives the best fit to the observed superlinearity (Fig.~\ref{fig:durint_scaling_model}a) and the distributions of $\tau$ and $\Delta \tau$ (Figs.~\ref{fig:scaling_diffalpha}--\ref{fig:interval_diff_resetprob} in SM). We verified the robustness of the results against moderate changes in $(\alpha, c_1, c_2)$ (Figs.~\ref{fig:scaling_diffalpha}--\ref{fig:interval_diff_resetprob}).
 
 We set $T=6500$ and discard the initial 5,000 simulation periods to eliminate the influence of the initial conditions. This leaves 1,500 effective simulation periods, which roughly correspond to 6 years of the empirical data.

%\showmatmethods % Display the Materials and Methods section

\section*{Acknowledgements}
T.K. acknowledges financial support from the Japan Society for the Promotion of Science Grants no. 15H05729 and 16K03551. The authors thank Naoki Masuda for useful comments on the manuscript and Research Project ``Network Science'' organized at International Institute for Advanced Studies for providing an opportunity to initiate the project.
\vspace{1cm}
%\showacknow % Display the acknowledgments section

% \pnasbreak splits and balances the columns before the references.
% If you see unexpected formatting errors, try commenting out this line
% as it can run into problems with floats and footnotes on the final page.
%\pnasbreak

% Bibliography
%\bibliography{emid_model}

%%%%%%%%%%%%%%%%%%%%%%%%%%%%%%%%%%%%
%%%%%%%% SI %%%%%%%%%%%%%%%%%%%%%%%%%%%

\clearpage

\setcounter{section}{0}
\setcounter{table}{0}
\setcounter{equation}{0}
\setcounter{figure}{0}
\setcounter{page}{1}
     
\renewcommand{\thetable}{S\arabic{table}}
\renewcommand{\thefigure}{S\arabic{figure}}
\renewcommand{\thesection}{S\arabic{section}}
\renewcommand{\theequation}{S\arabic{equation}}
\newcommand{\vect}[1]{\mbox{\boldmath $#1$}}

\begin{widetext}
\fontsize{16pt}{16pt}\selectfont
{\flushleft{{\textbf{Supplementary Materials:}}}  \\
\vspace{.3cm}
{\textbf{ ``Social dynamics of financial networks"}}} 

%\vspace{.2cm}
%\fontsize{12pt}{0pt}\selectfont
{\flushleft{\large Teruyoshi Kobayashi and Taro Takaguchi}}\\
\end{widetext}

%\fontsize{10.5pt}{13.5pt}\selectfont
\fontsize{9pt}{11pt}\selectfont

\section{Fitting procedure for the interval distribution}\label{sec:weibull_fit}

  As shown in Fig.~\ref{fig:durint_scaling_model}c, the empirical distribution of transaction interval $\Delta\tau$ for each bank pair does not follow a power law. We instead find that the interval distribution nicely fits a Weibull distribution for $1 \leq \Delta\tau < \Delta\tau_{\rm u}$, where $\Delta\tau_{\rm u}$ denotes a cutoff value. 

\begin{figure}[]
\begin{center}
    \includegraphics[width=.98\columnwidth]{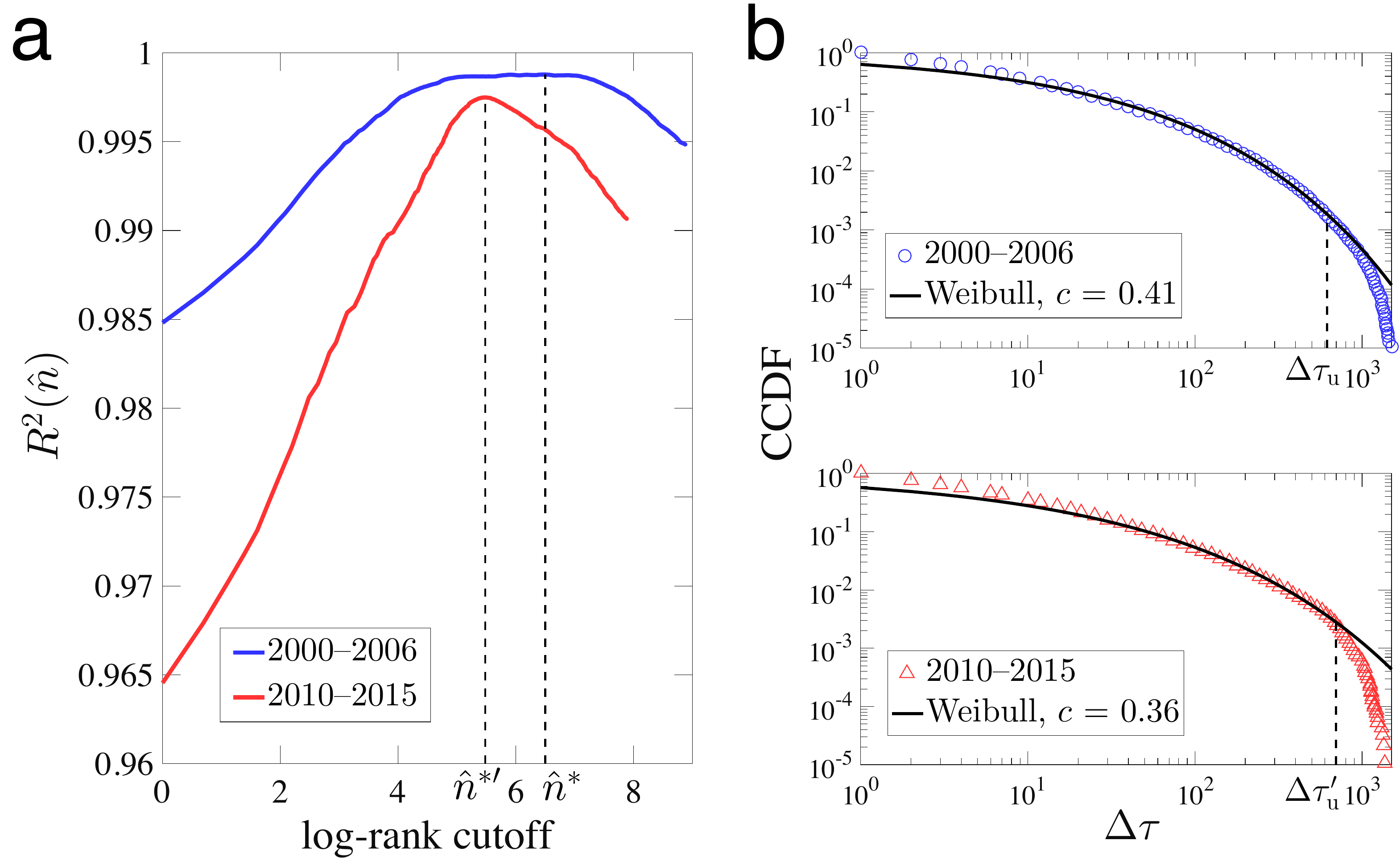}
    \end{center}
    \caption{Fitting of the interval distribution with a Weibull distribution. ({\textbf a}) Determination of the optimal log-rank cutoff $\hat{n}^*$. ({\textbf b}) Empirical CCDF of interval $\Delta\tau$ (symbols) and the Weibull distribution with the estimated parameters (lines). The cutoffs $\Delta\tau_u$  and $\Delta\tau_{u}^{\prime}$ are obtained from the optimal log-rank cutoffs $\hat{n}^*$ and $\hat{n}^{*\prime}$, respectively.}\label{fig:weibull_fit}
    \end{figure}

The complementary cumulative distribution function (CCDF) of a Weibull distribution~\cite{Weibull1951_SI} is given by
\begin{align}
  P_{c}(x) = \exp{\left\{ - \left( \frac{x}{\lambda}\right)^c\right\}}\; \text{ for $x>0$}, \label{eq:weibull_cdf}
\end{align}
where $c > 0$ and $\lambda>0$ are parameters. Distribution $P_c(x)$ can also be written as  $n_{x}/N_X$, where $N_X$ is the total number of interval values observed, and $n_x$ is the rank of interval length $x$ (i.e., $n_x$ is the number of observed interval values such that $\Delta \tau \geq x$). By taking the logarithm of $n_{x}/N_X = \exp{\left\{ - \left( x / \lambda \right)^c\right\}}$, we obtain the following expression~\cite{Sornette2006Book_SI}:
\begin{align}
(x_{n})^c = - \beta (\log{n_x}-\log{N_X}), \label{eq:rank_ols}
\end{align}
where $x_{n}$ represents the interval length whose rank is $n$ (i.e., $x_1 > x_2 > \ldots > x_{N_X})$, and $\beta$ is defined as $\beta \equiv \lambda^c$. 
We use Eq.~(\ref{eq:rank_ols}) to find $\beta$ and $c$ that give the best fit to a Weibull distribution.
We introduce $\hat{n}$, the logged rank of cutoff value $\Delta\tau_u$, and estimate parameters $(\beta, c)$ for a subset of the observed values of $\Delta\tau$, in a similar way as is done in the standard estimation procedure for a power-law exponent~\cite{Clauset2009Siam_SI}.
The cutoff $\Delta \tau_{\rm u}$ corresponds to the $e^{\hat{n}}$-th largest interval length.
Parameters $\beta$, $c$, and $\hat{n}$ are determined as follows.
\begin{enumerate}
\item For a given pair $(c, \hat{n})$, estimate $\beta$ in \eqref{eq:rank_ols} by the ordinary least squares (OLS). 
Repeat this for sufficiently many values of $c \in [0,1)$ (we set $c < 1$ because the tail of the empirical distribution of $\Delta \tau$ is apparently heavier than that of an exponential distribution). 
The estimate of $\beta$ is denoted by $\beta^*(c, \hat{n})$.

\item For $\hat{n}$ given in step 1,  find the optimal value of $c$, denoted by $c^*\left( \hat{n} \right)$, such that the coefficient of determination $R^2$ for the OLS regression is maximized, in which case $\beta = \beta^*(c^*(\hat{n}), \hat{n})$. Let $R^2(\hat{n})$ denote the maximum of $R^2$ for a given $\hat{n}$.

\item By repeating steps 1 and 2 for all the predefined values of $\hat{n}$, find the optimal cutoff value $\hat{n}^* \equiv \text{argmax}_{\hat{n}}{R^2(\hat{n})}$.
In the end, the estimates of the parameters are given by $\hat{n} = \hat{n}^*$, $c = c^*(\hat{n}^*)$, and $\beta = \beta^*(c^*(\hat{n}^*), \hat{n}^*)$.
\end{enumerate}
Figure~\ref{fig:weibull_fit}a illustrates the determination of the optimal log-rank cutoff $\hat{n}^*$. The inset of Fig.~\ref{fig:durint_scaling_model}c in the main text shows the OLS fit to \eqref{eq:rank_ols} when $\hat{n} = \hat{n}^*$ (note that $x_n$ corresponds to $\Delta\tau$ in that figure).  
Once $\hat{n}^*$ is determined, it is straightforward to obtain the corresponding cutoff $\Delta\tau_u$. Figure~\ref{fig:weibull_fit}b verifies the goodness of fit between the empirical CCDF and the estimated Weibull distribution.

 %%%%%%%%%%%%%%%%%%%%%%%%%%%%%%%%%%%%%% 

\section{Analytical solution for the fitness model with a finite size effect}\label{sec:analytical}

\subsection{\normalsize{Relationship between $N$ and $M$}} 

\ As we described in the main text, we assume that initially there are $\Np$ many isolated nodes.
Node~$i$ $(1 \leq i \leq \Np)$ is assigned a fitness $a_i \in [0,1]$ which is drawn from density $\rho(a)$.

The probability of edge formation between two nodes $i$ and $j$ is denoted by $u(a_i, a_j)$.
We define $N$ as the number of nodes connected with at least one edge and $M$ as the total number of edges in a network. We express $N$ and $M$ as functions of $\Np$:
\begin{align}
\begin{cases}
N &= (1- q_0(\Np)) \Np,\\
M &= \frac{\overline{k}(\Np) \Np}{2},
\end{cases}
\label{eq:NM_SI}
\end{align}
where $q_0(\Np)$ is the probability of a randomly chosen node being isolated (i.e., no edges attached) and $\overline{k}(\Np)$ is the average degree over all nodes including isolated ones.
Thus, to obtain the functional forms of $N$ and $M$, we need to get the functional forms of $q_0(\Np)$ and $\overline{k}(\Np)$.
In the following, we first derive the functional forms of $q_0(\Np)$ and $\overline{k}(\Np)$ in a general setting.
Then, we will restrict our attention to the case with $\rho(a) = 1$ (i.e., a uniform distribution) and $u(a_i, a_j) = (a_i a_j)^\alpha$ to explain the empirical superlinear relation between $N$ and $M$ in the same specification as in the main text. 

%%%%%%%%%%%%%%%%%%%%%%%%%%%%%%%%%%%%%%%%%%
\subsection{\normalsize{Average degree of networks including isolated nodes, $\overline{k}(\Np)$}} 

\ Given the fitnesses of all nodes $\vect{a} = (a_1, a_2, \ldots, a_{\Np})$, the probability that node $i$ has degree $k_i$ is
\begin{align}
g(k_i | \vect{a}) &= \sum_{\vect{c}_i} \left[ \prod_{j \neq i} u(a_i, a_j)^{c_{ij}} (1- u(a_i, a_j))^{1-c_{ij}} \right] \notag \\ 
&\times \delta\left(\sum_{j \neq i} c_{ij}, k_i \right),
\label{eq:g_ki}
\end{align}
where $c_{ij} \in \left\{ 0, 1\right\}$ is the $(i,j)$-element of the adjacency matrix and $\vect{c}_i = (c_{1i}, c_{2i}, \ldots, c_{\Np i})^\top$ is the $i$th column vector.
Function $\delta(x, y)$ denotes the Kronecker delta.
Let us redefine a product term in the square bracket of \eqref{eq:g_ki} as
\begin{align}
f_j(c_{ij}; a_i, a_j) &\equiv u(a_i, a_j)^{c_{ij}} (1- u(a_i, a_j))^{1-c_{ij}}.
\label{eq:def_f}
\end{align}
Since $g(k_i | \vect{a})$ is the convolution of $\left\{ f_j(c_{ij}; a_i, a_j) \right\}_j$, its generating function
\begin{align}
\hat{g}_i(z | \vect{a}) \equiv \sum_{k_i} z^{k_i} g(k_i | \vect{a})
\end{align}
is decomposed as
\begin{align}
\hat{g}_i(z | \vect{a}) = \prod_{j \neq i} \hat{f}_j(z; a_i, a_j),
\end{align}
where $\hat{f}_j$ is the generating function of $f_j(c_{ij}; a_i, a_j)$, given by
\begin{align}
 \hat{f}_j(z; a_i, a_j) \equiv \sum_{a_{ij}} z^{a_{ij}} f_j(a_{ij}; a_i, a_j).
 \label{eq:def_fhat}
\end{align}

Degree distribution $p(k_i; \Np)$ is defined by the probability that node~$i$ has degree $k_i$ and is related to $g(k_i | \vect{a})$ such that
\begin{align}
p(k_i ; \Np) = \int g(k_i | \vect{a}) \rho(\vect{a}) d \vect{a},
\label{eq:p_ki}
\end{align}
where we define $\rho(\vect{a}) \equiv \prod_i \rho(a_i)$ and $d \vect{a} \equiv \prod_i da_i$. Therefore, differentiation of $\hat{g}_i(z | \vect{a})$ with respect to $z$ gives the average degree $\overline{k}(\Np)$:
\begin{align}
\overline{k}(\Np) &= \sum_{k_i} k_i p(k_i; \Np) \nonumber\\
&= \sum_{k_i} k_i \int g(k_i | \vect{a})\rho(\vect{a}) d\vect{a} \nonumber\\
&= \frac{d}{dz} \int \hat{g}_i(z | \vect{a}) \rho(\vect{a}) d \vect{a} \Bigr|_{z=1} \nonumber\\
&= \frac{d}{dz} \int \rho(a_i) da_i \prod_{j \neq i} \int \hat{f}_j(z; a_i, a_j) \rho(a_j) da_j \Bigr|_{z=1} \nonumber\\
&= \int \rho(a_i) da_i \frac{d}{dz} \left[ \int \hat{f}(z; a_i, h) \rho(h) dh \right]^{\Np-1} \Bigr|_{z=1} \nonumber\\
&= (\Np-1)  \int \rho(a_i) d a_i \left[ \int d a \rho(a) \hat{f}(z; a_i, a) \right]^{\Np-2} \notag \\
&\hspace{1cm} \times \int d a \rho(a) \frac{d}{dz} \hat{f}(z; a_i, a)\Bigr|_{z=1}.
\label{eq:k1}
\end{align}
From Eqs.~(\ref{eq:def_f}) and (\ref{eq:def_fhat}), we have $\hat{f}(z; a_i, a) = \sum_{c_{ij}} z^{c_{ij}} f(c_{ij}; a_i, a) = (z-1)u(a_i, a) + 1$. It follows that  
\begin{align}
&\int da \rho(a) \hat{f}(z; a_i, a) = (z-1) \int da \rho(a) u(a_i, a) + 1,\\
&\int da \rho(a) \frac{d}{dz} \hat{f}(z; a_i, a) = \int da \rho(a) u(a_i, a).
\end{align}
Substituting these into Eq.~(\ref{eq:k1}) leads to
\begin{align}
\overline{k}(\Np) = (\Np-1) \int \int d a d a^\prime \rho(a)  \rho(a^\prime) u(a, a^\prime).
\label{eq:k_avg}
\end{align}
It should be noted that \eqref{eq:k_avg} is equivalent to Eq.~(21) of Ref.~\cite{Boguna2003PRE_SI}.

%%%%%%%%%%%%%%%%%%%%%%%%%%%%%%%%%
\subsection{\normalsize{Probability of node isolation, $q_0(\Np)$}}

\ From \eqref{eq:p_ki}, the probability of a node being isolated, $q_0(\Np) \equiv p(k_i = 0; \Np)$, is given by 
\begin{align}
q_0(\Np) 
&= \int g(k_i =0 | \vect{a})\rho(\vect{a}) d \vect{a} \nonumber\\
%
%&= \int d a_i \rho(a_i) \left[ \int (1- u(a_i, a))  \rho(a) d a \right]^{\Np-1} \nonumber\\
%
&= \int d a_i \rho(a_i) \left[ 1 - \int u(a_i, a) \rho(a) d a \right]^{\Np-1}.
\label{eq:q_0}
\end{align}

%%%%%%%%%%%%%%%%%%%%%%%%%%%%%%%%%%%%%%%%
\subsection{\normalsize{Special case: $\rho(a) = 1$ and $u(a, a^\prime) = (a  a^\prime)^\alpha$}}

\ Substituting $\rho(a) = 1$ and $u(a, a^\prime) = (a a^\prime)^\alpha$ into Eq.~(\ref{eq:k_avg}) gives
\begin{align}
\overline{k}(\Np) %&= (\Np-1) \int \int d a d a^\prime  (a_i a_j)^\alpha \nonumber\\
&=  \left( \frac{1}{\alpha + 1}\right)^2 (\Np-1).
\end{align}
Similarly, substituting the same conditions into Eq.~(\ref{eq:q_0}) gives
\begin{align}
q_0(\Np) &= \int d a_i  \left( 1 - \frac{1}{\alpha+1} a_i^\alpha \right)^{\Np-1}.
\end{align}
By rewriting the integrand as  $x = 1 - \frac{1}{\alpha+1} a_i^\alpha$, we have  
\begin{align}
q_0(\Np) &= \frac{1}{\alpha} (\alpha + 1)^{\frac{1}{\alpha}} \int_{1-\frac{1}{\alpha+1}}^1 (1-x)^{\frac{1}{\alpha}-1} x^{\Np-1} dx  \notag \\
%
%&= \frac{1}{\alpha} (\alpha + 1)^{\frac{1}{\alpha}} \left[ \int_{0}^1 (1-x)^{\frac{1}{\alpha}-1} x^{\Np-1} dx \right. \notag \\
 %&\left. \hspace{1cm} - \int_{0}^{1-\frac{1}{\alpha+1}} (1-x)^{\frac{1}{\alpha}-1} x^{\Np-1} dx\right] \notag \\
%
&= \frac{1}{\alpha} (\alpha + 1)^{\frac{1}{\alpha}} \left[ {\rm B}\left( \Np, \frac{1}{\alpha}\right) - {\rm B}_{1-\frac{1}{\alpha+1}} \left( \Np, \frac{1}{\alpha}\right)\right],
\end{align}
where ${\rm B}(x,y) \equiv \int_0^1 t^{x-1} (1-t)^{y-1} dt$ is the beta function and ${\rm B}_z(x,y) \equiv \int_0^z t^{x-1} (1-t)^{y-1} dt$ ($0 < {\rm Re}(z) \leq 1$) is the incomplete beta function.
Combining these results with Eq.~(\ref{eq:NM_SI}), we end up with
\begin{align}
\begin{cases}
N &= \Np \left[ 1- \frac{1}{\alpha} (\alpha + 1)^{\frac{1}{\alpha}} \left( {\rm B}\left( \Np, \frac{1}{\alpha}\right) - {\rm B}_{1-\frac{1}{\alpha+1}} \left( \Np, \frac{1}{\alpha}\right)\right) \right],\\
M &= \left( \frac{1}{\alpha + 1} \right)^2 \frac{\Np(\Np-1)}{2}.
\end{cases}
\end{align}
If $\Np$ is sufficiently large, then $q_0(\Np) \simeq 0$ and thereby $N \simeq \Np$ and $M \simeq (1/\alpha +1)^2 N(N-1)/2 \propto N^2$. Therefore, 
the solution is consistent with that of the previous studies~\cite{Caldarelli2002PRL_SI,Boguna2003PRE_SI,DeMasi2006PRE_SI} in the absence of the finite-size effect.
 
%\clearpage

  \vspace{1cm}

  \section{Dynamics of weights}
  
\subsection{\normalsize{Empirical observation}}

\ On top of the edge dynamics that we discussed in the main text, the dynamics of edge weights also exhibits specific patterns.
Let us define the weight of an edge, $w_{ij,t}$, as the total amount of funds transferred from bank $i$ to $j$ on day $t$.
We define the growth rate of edge weights as $r_{ij,t} \equiv \log{(w_{ij,t+1}/w_{ij,t})}$ for bank pair $(i, j)$ such that $w_{ij,t+1}w_{ij,t}>0$~\cite{Gautreau2009PNAS_SI}.
The distribution of $r_{ij,t}$, aggregated over all pairs and all $t$, exhibits a symmetric triangular shape with a distinct peak at 0 (Fig.~\ref{fig:weightgrowth}a). The shape of the distribution indicates that a large fraction of bank pairs do not change the amount of funds when they keep trading, and if they change the amount, the rate of change will be typically small.
A similar sort of triangular-shaped distribution of the growth rate of weights has also been found in networks of email exchanges~\cite{Godoy2016Plosone_SI}, airlines~\cite{Gautreau2009PNAS_SI} and cattle trades between stock farming facilities~\cite{Bajardi2011PlosoneCattle_SI}.

\begin{figure*}[]
\begin{center}
    \includegraphics[width=1.5\columnwidth]{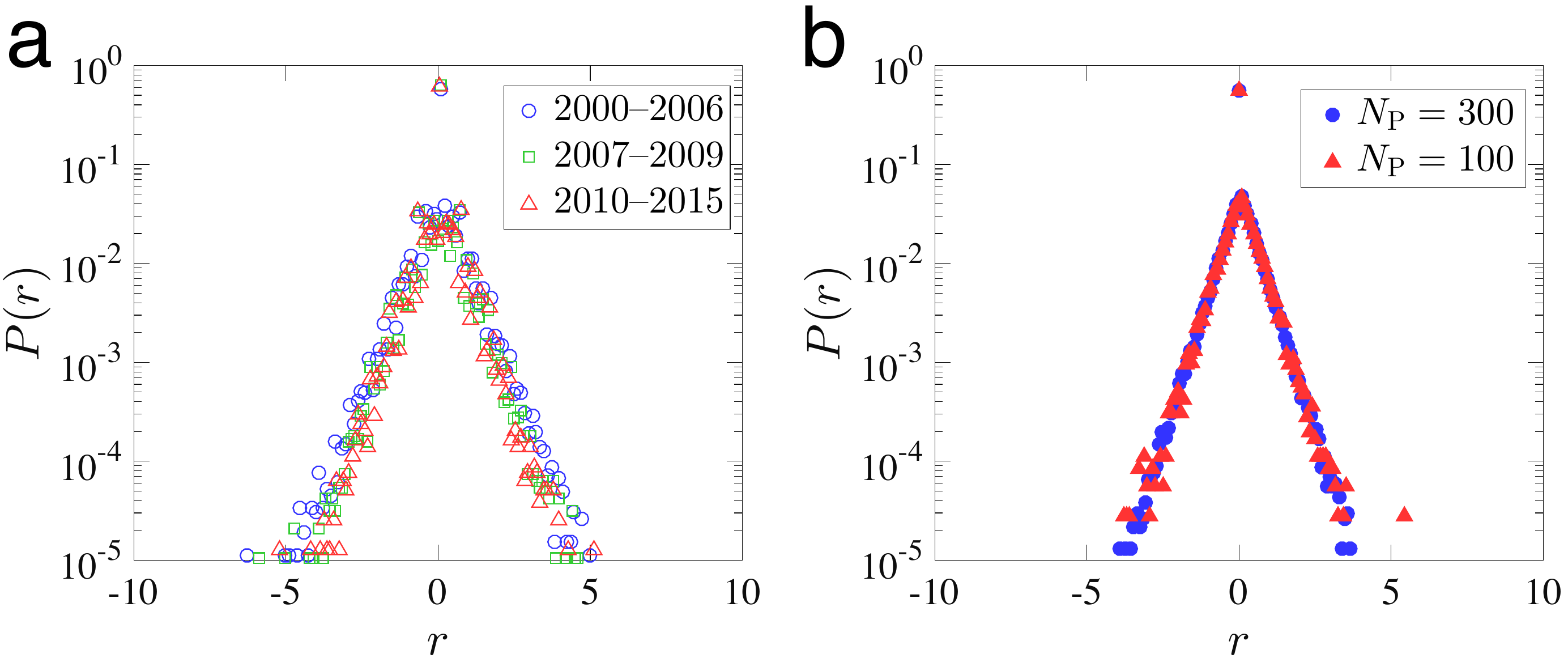}
    \end{center}
    \caption{Distribution of the growth rates of edge weights $r$ (aggregated over all bank pairs) for ({\textbf a}) the data and ({\textbf b}) the model.}\label{fig:weightgrowth}
    \end{figure*}

\begin{figure*}[]
\begin{center}
     \includegraphics[width=1.5\columnwidth]{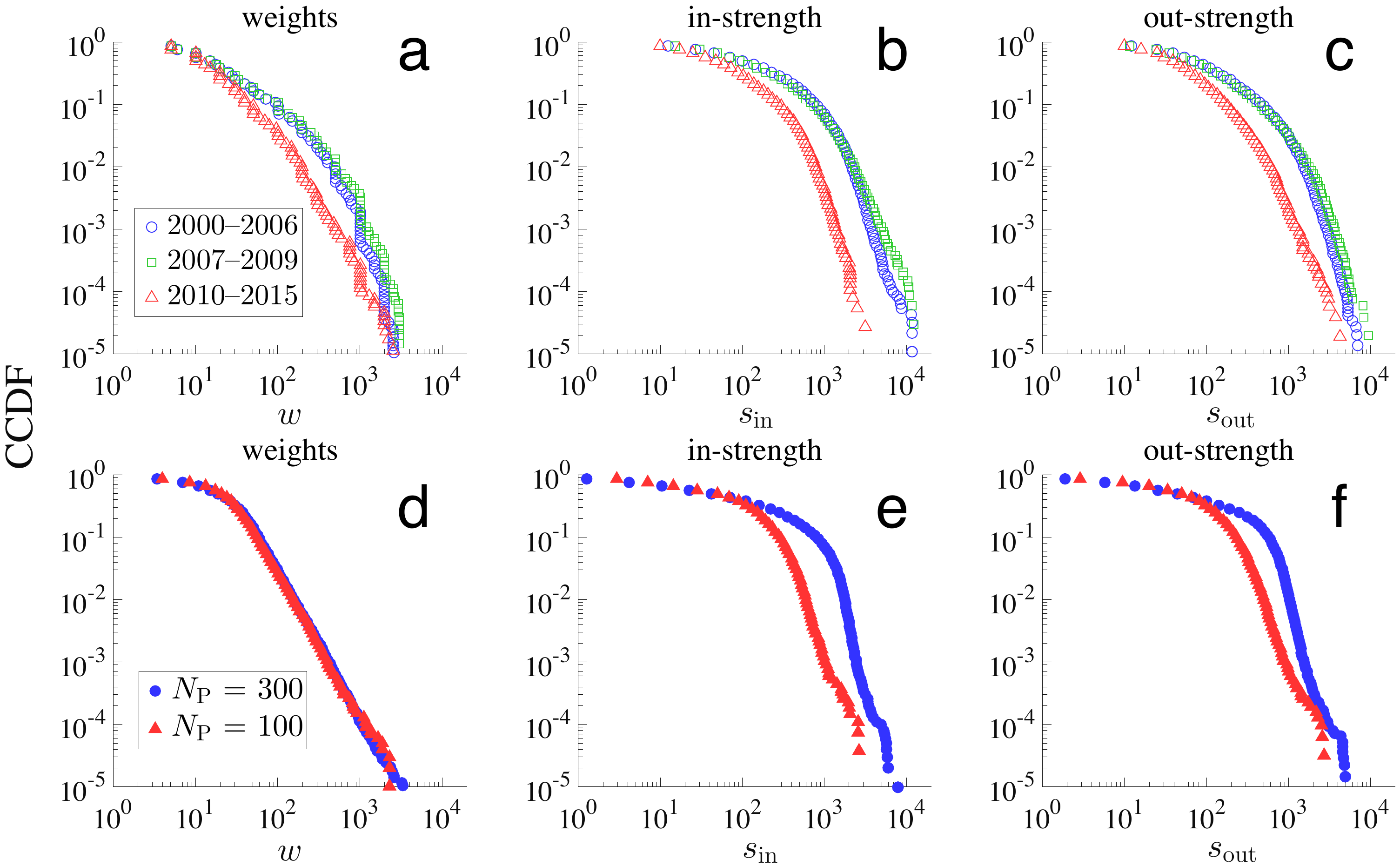}
    \end{center}
    \caption{CCDF of edge weights and strength. The in- and out-strength of bank $i$ are defined as 
$s_{{\rm in},i} = \sum_{j\neq i}w_{ji}$ and 
$s_{{\rm out},i} = \sum_{j\neq i}w_{ij}$, respectively. ({\textbf a})--({\textbf c}) The data. ({\textbf d})--({\textbf f}) The model.\label{fig:weight_dist}}
\end{figure*}

\begin{figure*}
\begin{center}
     \includegraphics[width=1.5\columnwidth]{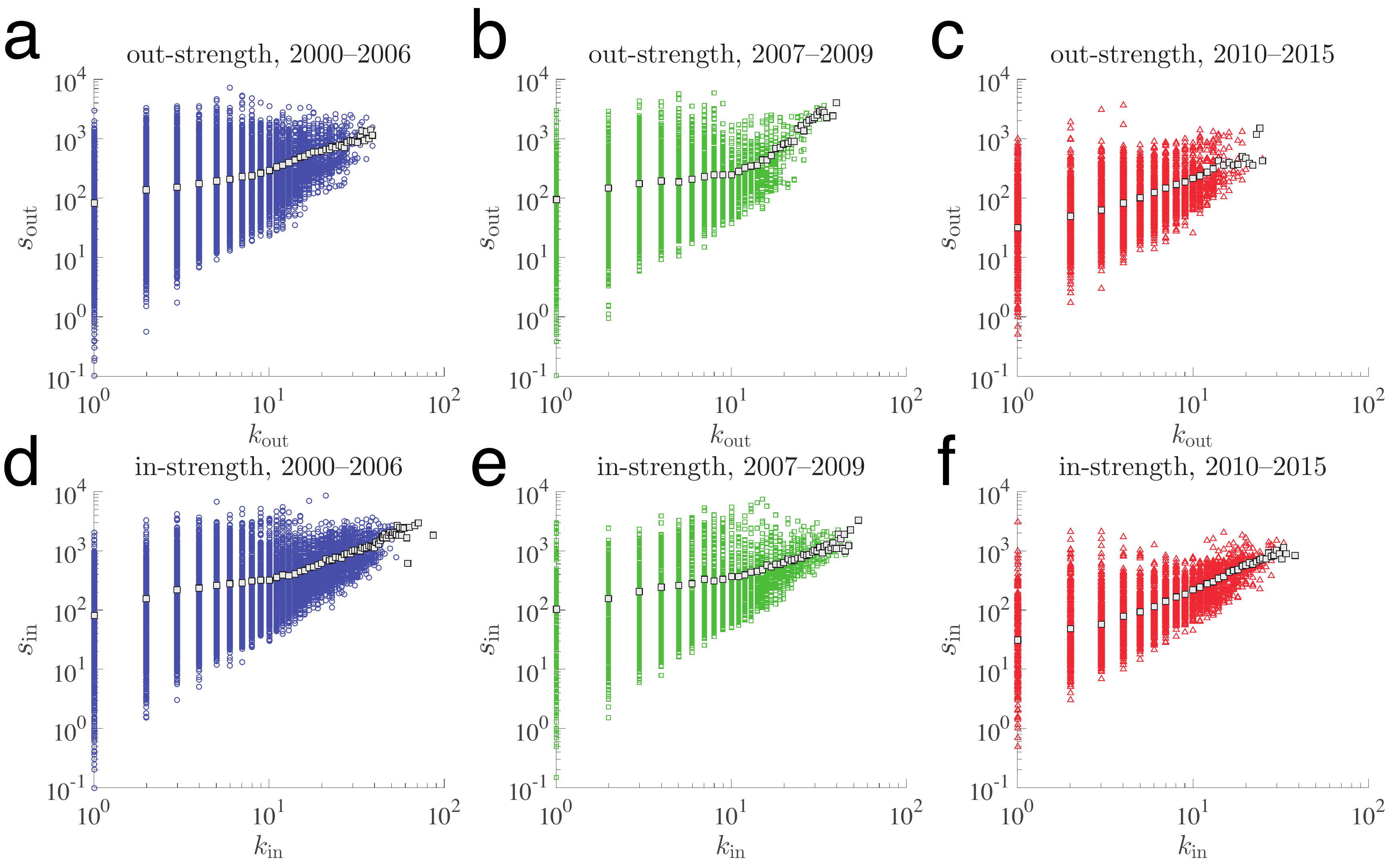}
    \end{center}
    \caption{Scatter plots of strength against degree in the empirical data. Light-gray square denotes the average strength for a given degree. 80\% of the markers are randomly removed since most of them are overlapped. See Fig.~\ref{fig:weight_dist} for the definition of the strength.}\label{fig:strength_data}
\end{figure*}

\begin{figure*}
\begin{center}
     \includegraphics[width=1.35\columnwidth]{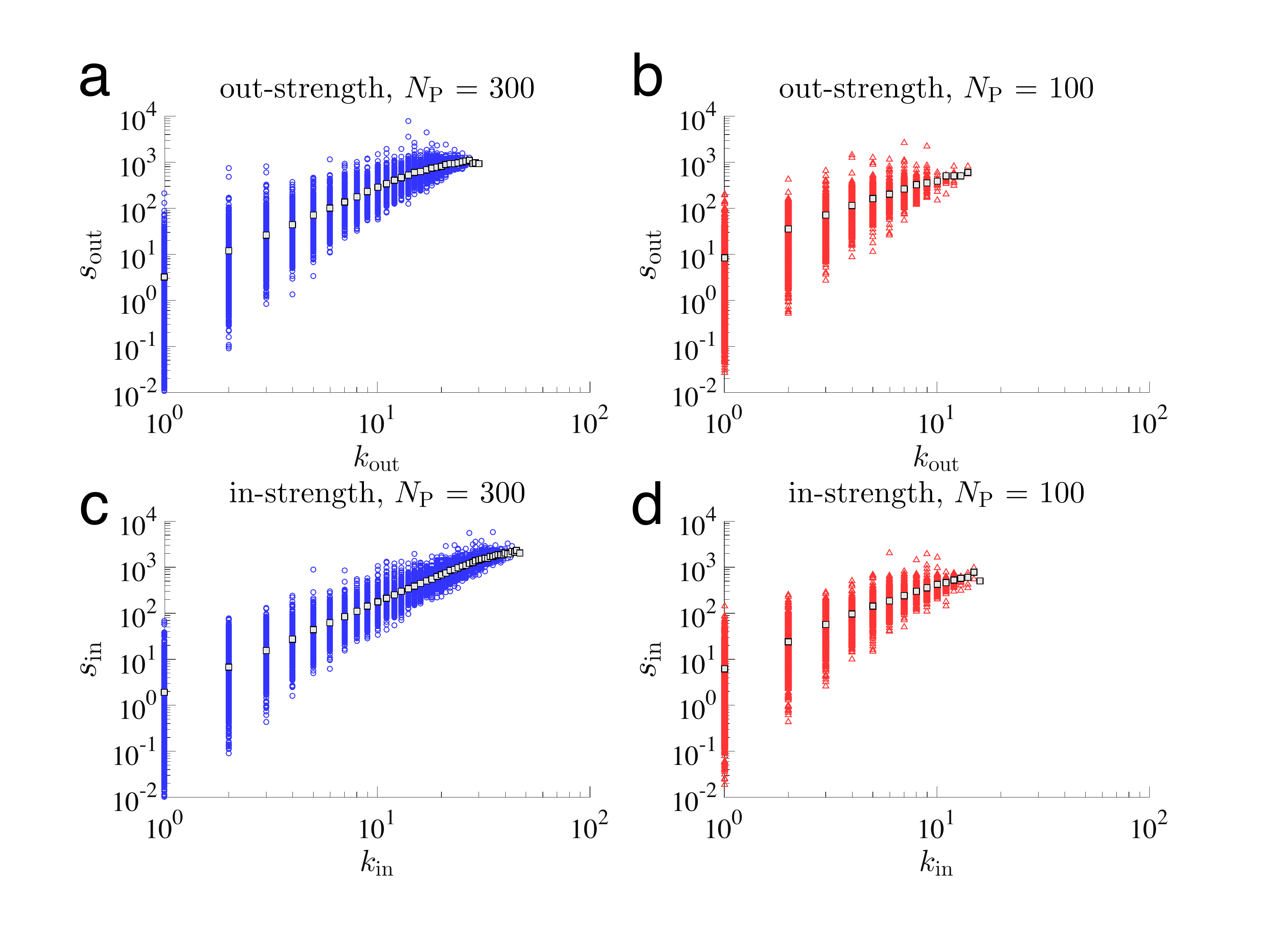}
    \end{center}
    \caption{Scatter plots of strength against degree in the model. Light-gray square denotes the average strength for a given degree.  80\% of the markers are randomly removed as in Fig.~\ref{fig:strength_data}. See Fig.~\ref{fig:weight_dist} for the definition of the strength.}\label{fig:strength_model}
\end{figure*}

\subsection{\normalsize{Model of weight dynamics}}

\ To reproduce the dynamics of edge weights, we add the following step to the model. Let us consider the edge between $i$ and $j$ formed in day $t$.  If there is an edge from $i$ to $j$ in day $t-1$, then the edge weights in day $t$ is given by
\begin{align}
  w_{ij,t} \equiv
  \begin{cases}
   w_{ij,t-1}  & \text{with probability $1-q$}, \\
   \kappa\nu_{ij,t}p_{ij,t} & \text{with probability $q$},
  \end{cases}
\end{align}
where random variable $\nu_{ij,t}$ takes different values across bank pairs and are assumed to follow a power-law distribution with exponent $\eta$ to maximize the fit to $P(r)$ (Fig.~\ref{fig:weightgrowth}) and the empirical weight distribution (Fig.~\ref{fig:weight_dist} a--c). 
Positive constant $\kappa$ is introduced to match the scale of edge weights with that of the data (i.e., millions of euros). 
On the other hand, if there is no edge from $i$ to $j$ in day $t-1$ but there is in day $t$, 
then
 \begin{align}
  w_{ij,t} \equiv
   \kappa\nu_{ij,t}p_{ij,t}.
\end{align}
Any non-adjacent pairs $(i,j)$ has $w_{ij,t}=0$.

We set the weight parameters as $(q, \kappa, \eta) =(0.5, 80, 3.3)$ to fit $P(r)$ and the simulated weight distributions with the empirical ones, respectively.
Figures~\ref{fig:weightgrowth}b and \ref{fig:weight_dist}d--f show that our model of weight dynamics successfully replicates the empirical distributions.

%\end{widetext}

     %[do not remove: created by weibull_logrank.m]

   %[do not remove: created by Nbar_Plot_durint.m]

\newpage

\begin{figure*}
\begin{center}
    \includegraphics[width=1.35\columnwidth]{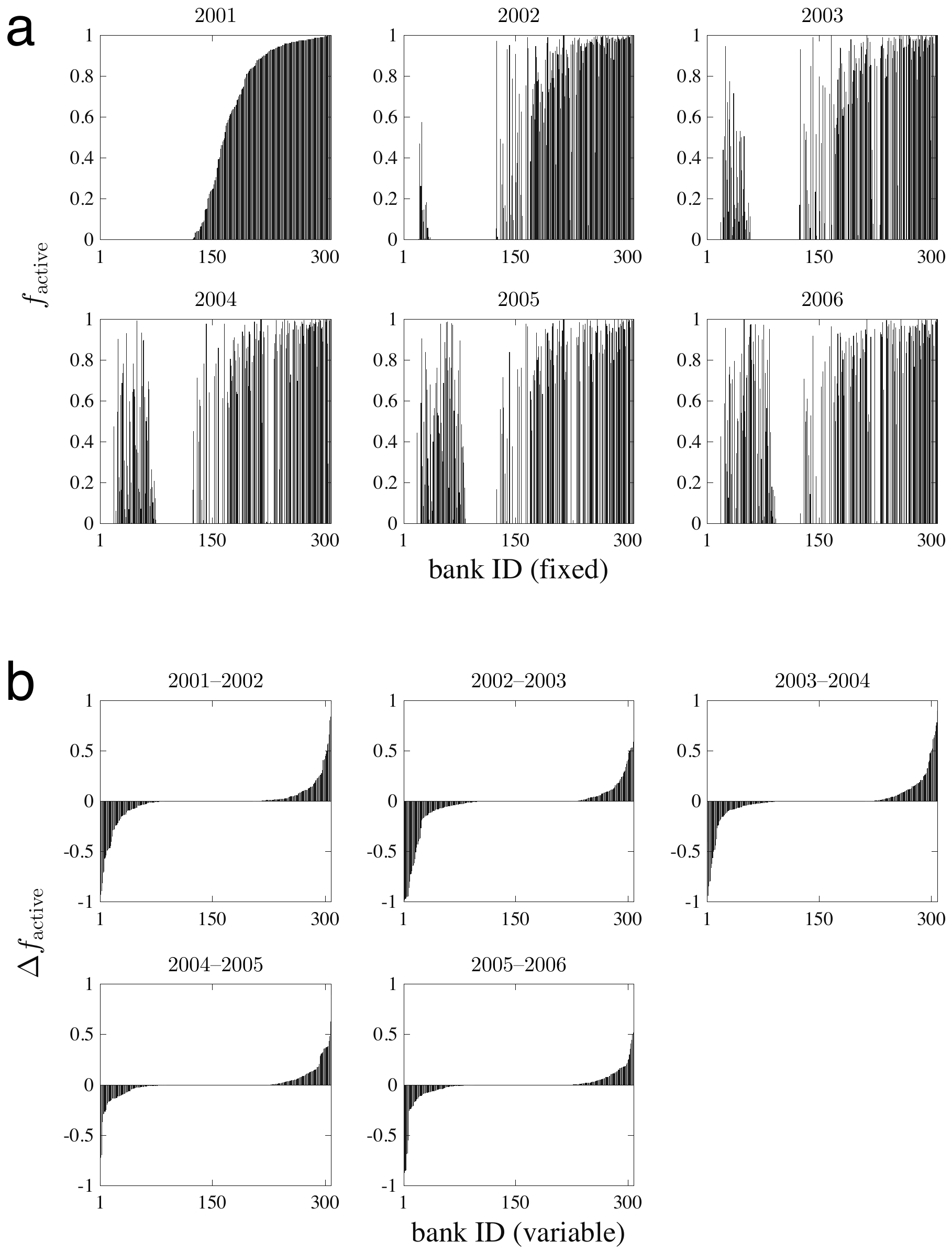}
    \end{center}
    \caption{Visualization of the annual turnover in the set of active banks. ({\textbf a}) Fraction of active days for each bank, denoted by $f_{\rm active}$. Bank IDs are sorted and fixed in the ascending order of $f_{\rm active}$ in the year of 2001.  ({\textbf b}) Changes in $f_{\rm active}$ of each bank between two consecutive years, denoted by $\Delta f_{\rm active}$. Bank IDs are sorted in the ascending order of $\Delta f_{\rm active}$ in each panel. ({\textbf c}) Simulated $f_{\rm active}$ in the model with $\Np =300$. $y_i$ denotes ``year $i$", where each ``year" consists of 250 simulation periods. Bank IDs are sorted and fixed in the ascending order of $f_{\rm active}$ in the ``year" of $y_1$. 
  ({\textbf d}) $\Delta f_{\rm active}$ corresponding to the simulated $f_{\rm active}$. Bank IDs are sorted in the ascending order of $\Delta f_{\rm active}$ in each panel.}   
  \label{fig:turnover}
    \end{figure*}%[do not remove: created by TradeProperty.m]
  
     \newpage
     
     \begin{figure*}
\begin{center}
    \includegraphics[width=1.35\columnwidth]{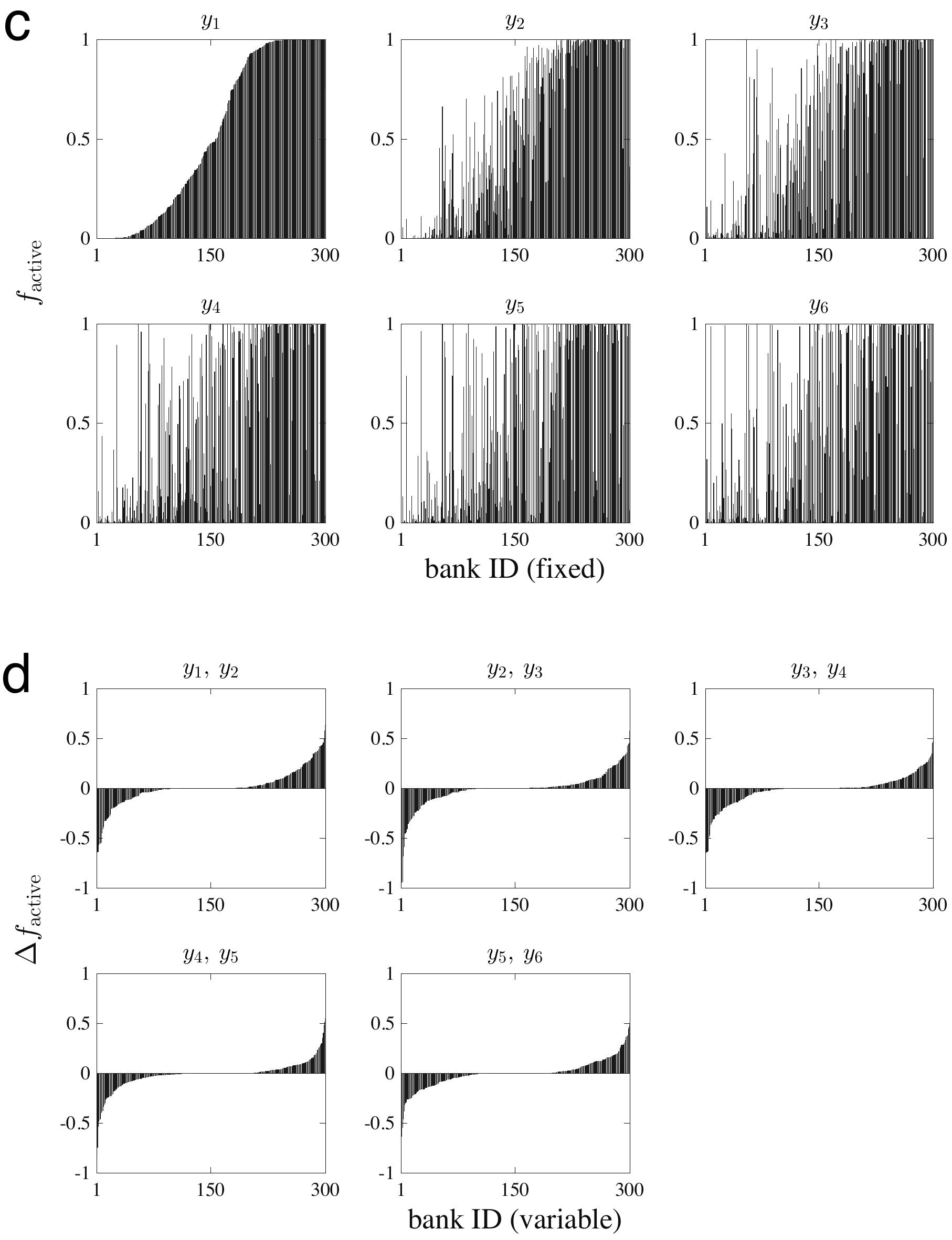}
    \end{center}
    \end{figure*}%[do not remove: created by Nbar_model.m]

     \newpage 
     
 \begin{figure*}
\begin{center}
    \includegraphics[width=1.5\columnwidth]{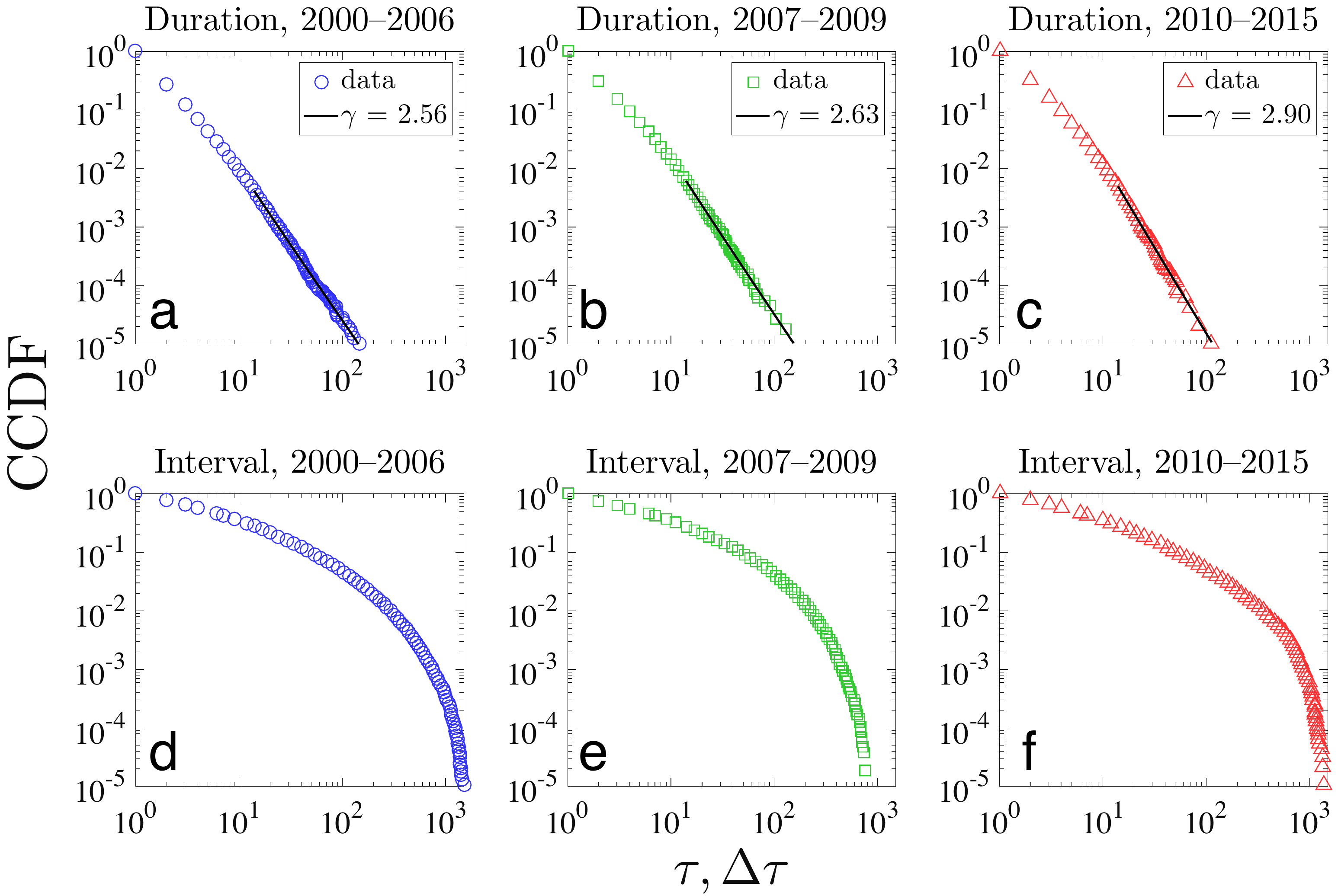}
    \end{center}
    \caption{CCDF of duration $\tau$ and interval $\Delta \tau$ for transactions of each bank pair, aggregated over all pairs. The exponents are estimated by using the Matlab codes downloaded from~\cite{ClausetHP_SI}
    , which is based on~\cite{Clauset2009Siam_SI}
    .}\label{fig:durint_data}
    \end{figure*}%[do not remove: created by TradeProperty.m]

\begin{figure*}
\begin{center}
     \includegraphics[width=1.5\columnwidth]{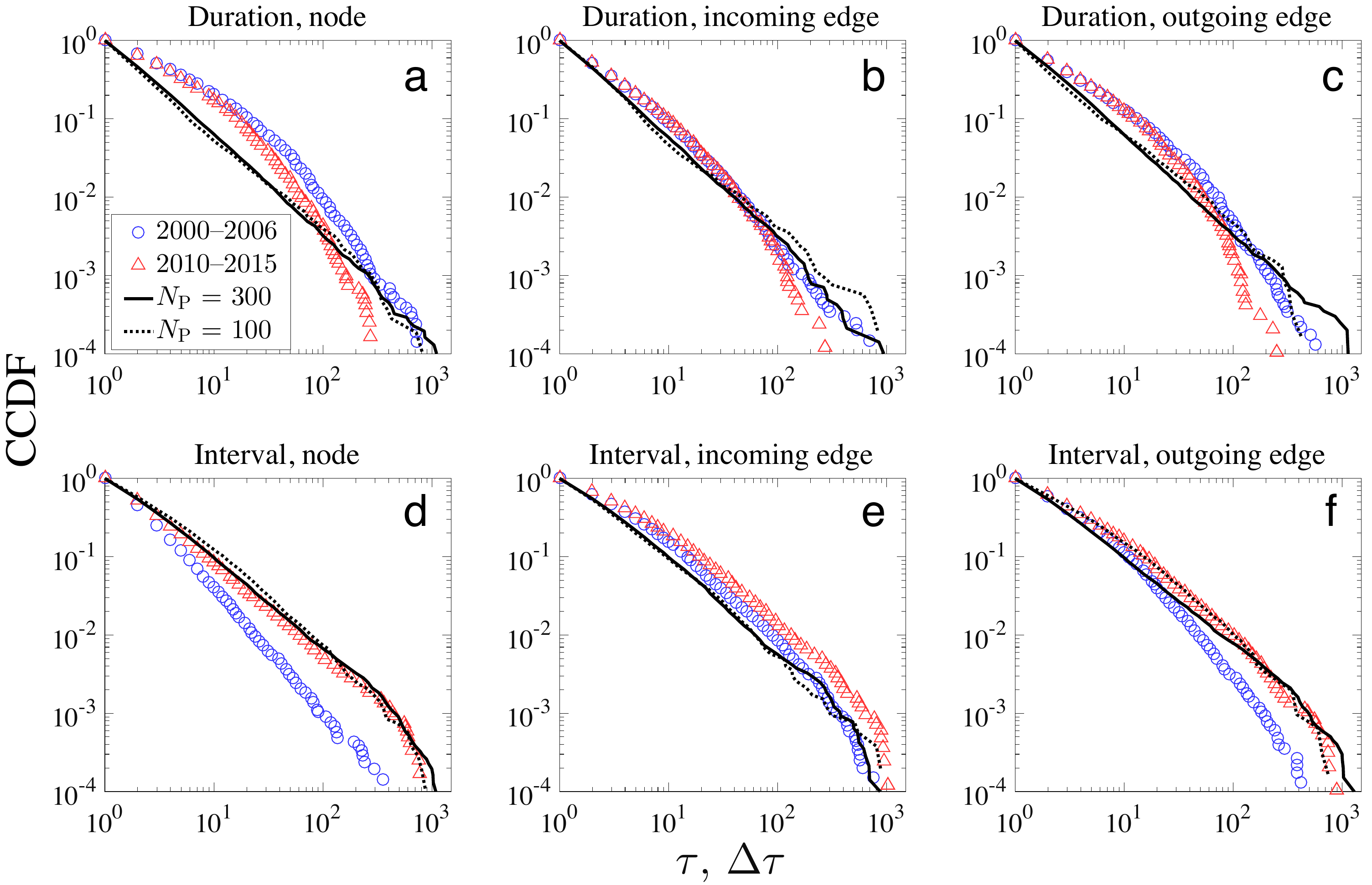}
    \end{center}
    \caption{CCDF of duration $\tau$ and interval $\Delta \tau$ redefined for node activity. The upper panels show CCDF of $\tau$, defined by the consecutive trading days on each of which ({\textbf a}) a node is active; ({\textbf b}) a node has at least one incoming edge; ({\textbf c}) a node has at least one outgoing edge. The lower panels show CCDF of $\Delta\tau$, defined by the number of interval days during which ({\textbf d}) a node is inactive; ({\textbf e}) a node has no incoming edge; ({\textbf f}) a node has no outgoing edge.}\label{fig:durint_model_node_link}
\end{figure*}%[do not remove: created by Nbar_Plot_durint.m]

  \newpage 
  
\begin{figure*}
\begin{center}
        \includegraphics[width=1.5\columnwidth]{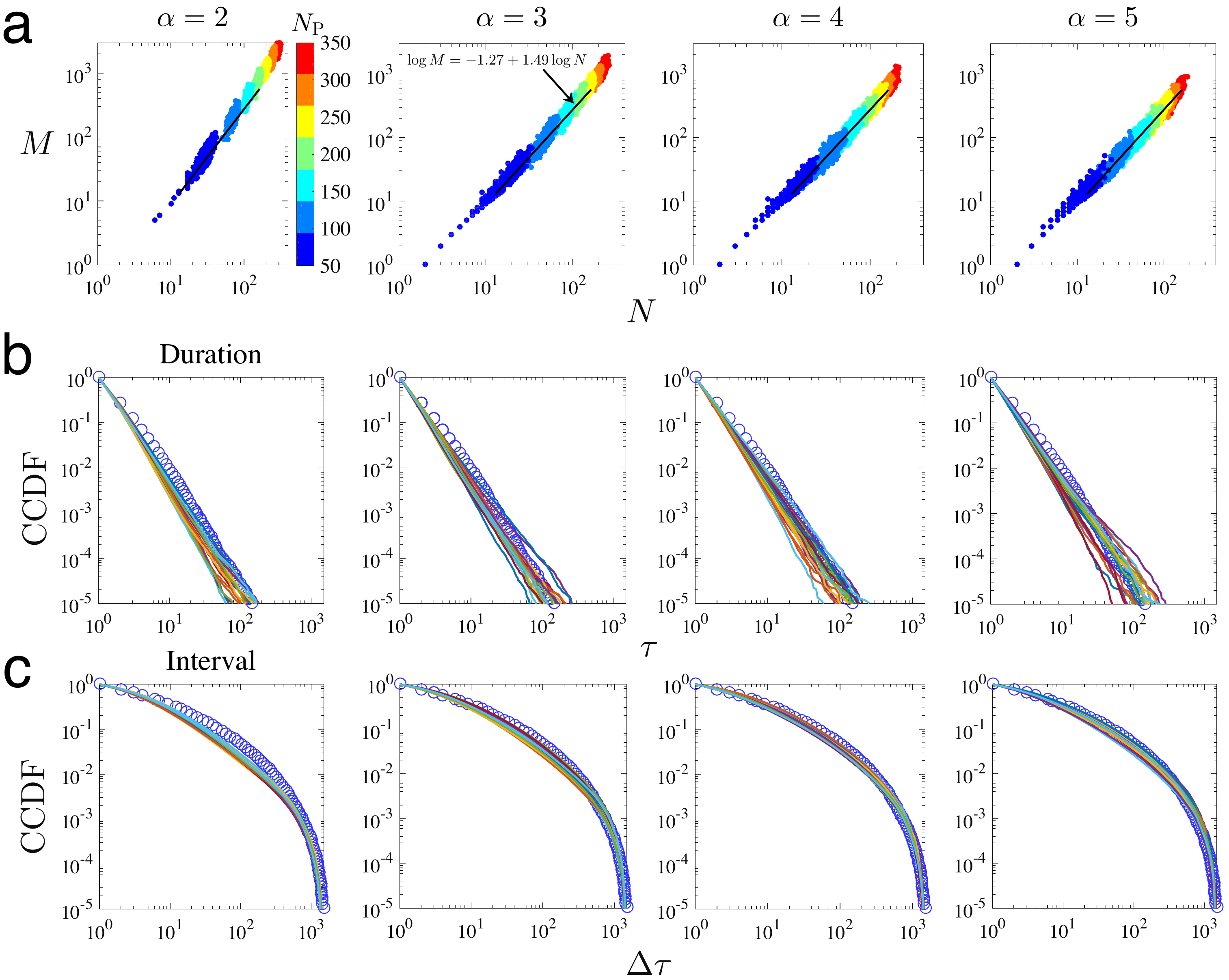}
    \end{center}
    \caption{Effect of varying parameter $\alpha$ on scaling relations. In ({\textbf a})--({\textbf c}), each column illustrates the simulation results under a fixed value of $\alpha$ annotated above panel a. ({\textbf a}) Scaling of $M$ with $N$. Solid line indicates the regression line of the empirical scaling. Model networks are generated 500 times for a given $\Np$. ({\textbf b}) Empirical and simulated CCDF of transaction duration $\tau$ for bank pairs. ({\textbf c}) CCDF of transaction interval $\Delta \tau$ for bank pairs. In ({\textbf b}) and ({\textbf c}), blue circle denotes the empirical CCDF for the 2000--2006 period. To visualize the stability of simulated CCDF, 20 lines of CCDF generated by independent 20 runs are plotted for $\Np=300$.}\label{fig:scaling_diffalpha}
 %[do not remove: created by Nbar_varyingstate_model.m]
\end{figure*}

\begin{figure*}
\begin{center}
     \includegraphics[width=1.5\columnwidth]{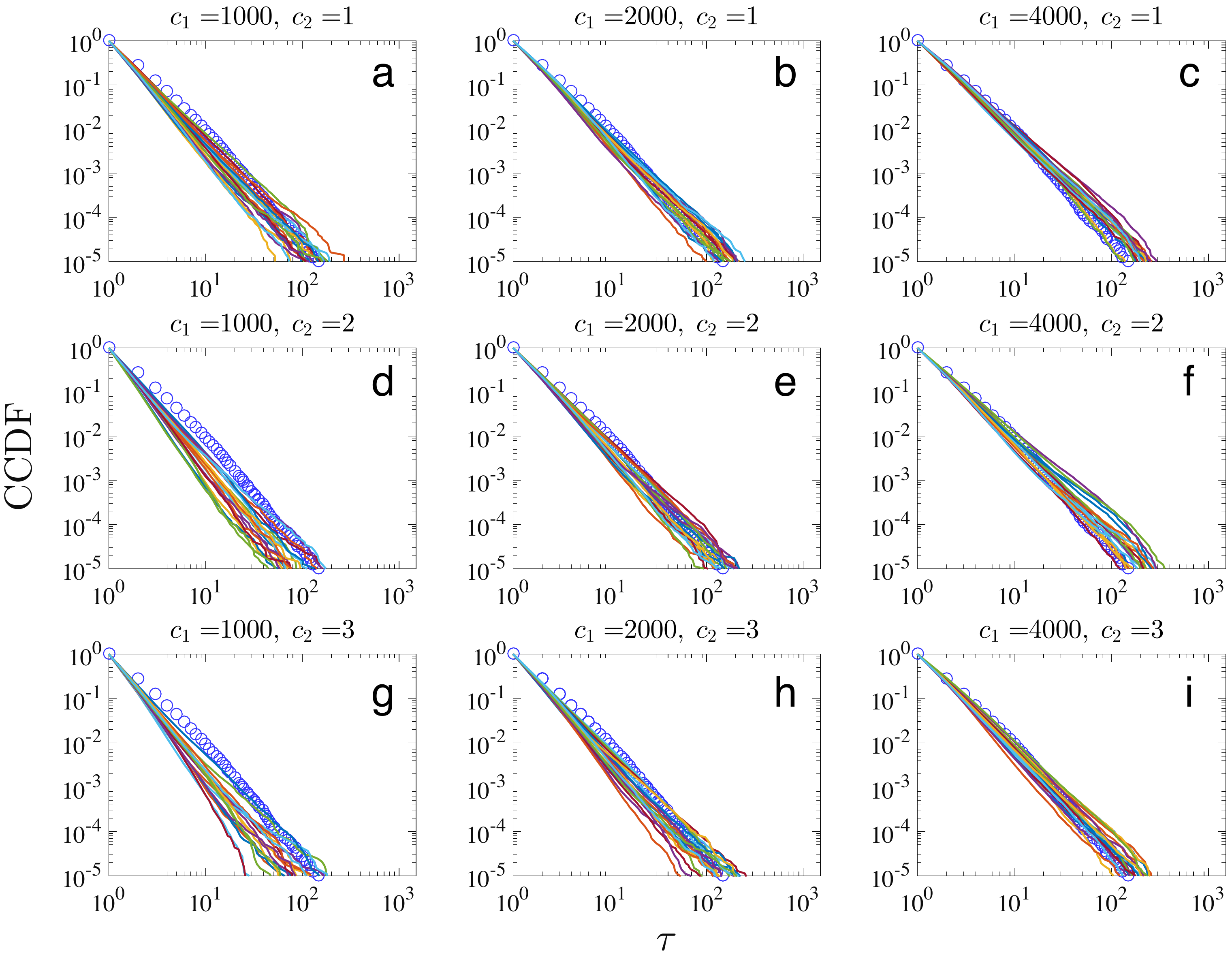}
    \end{center}
    \caption{Effect of altering the specification of reset probability $h(a)$ on the duration distribution of pairwise transactions. Parameters $c_1$ and $c_2$ are defined by $h(a) = c_{1}^{-1}a^{c_2}$. Blue circle denotes the empirical CCDF for the 2000--2006 period. 20 lines of CCDF generated by independent 20 runs are plotted for $\Np=300$.}\label{fig:duration_diff_resetprob}%[do not remove: created by Nbar_durint_diff_resetprob.m]
\end{figure*}

 \begin{figure*}
\begin{center}
     \includegraphics[width=1.5\columnwidth]{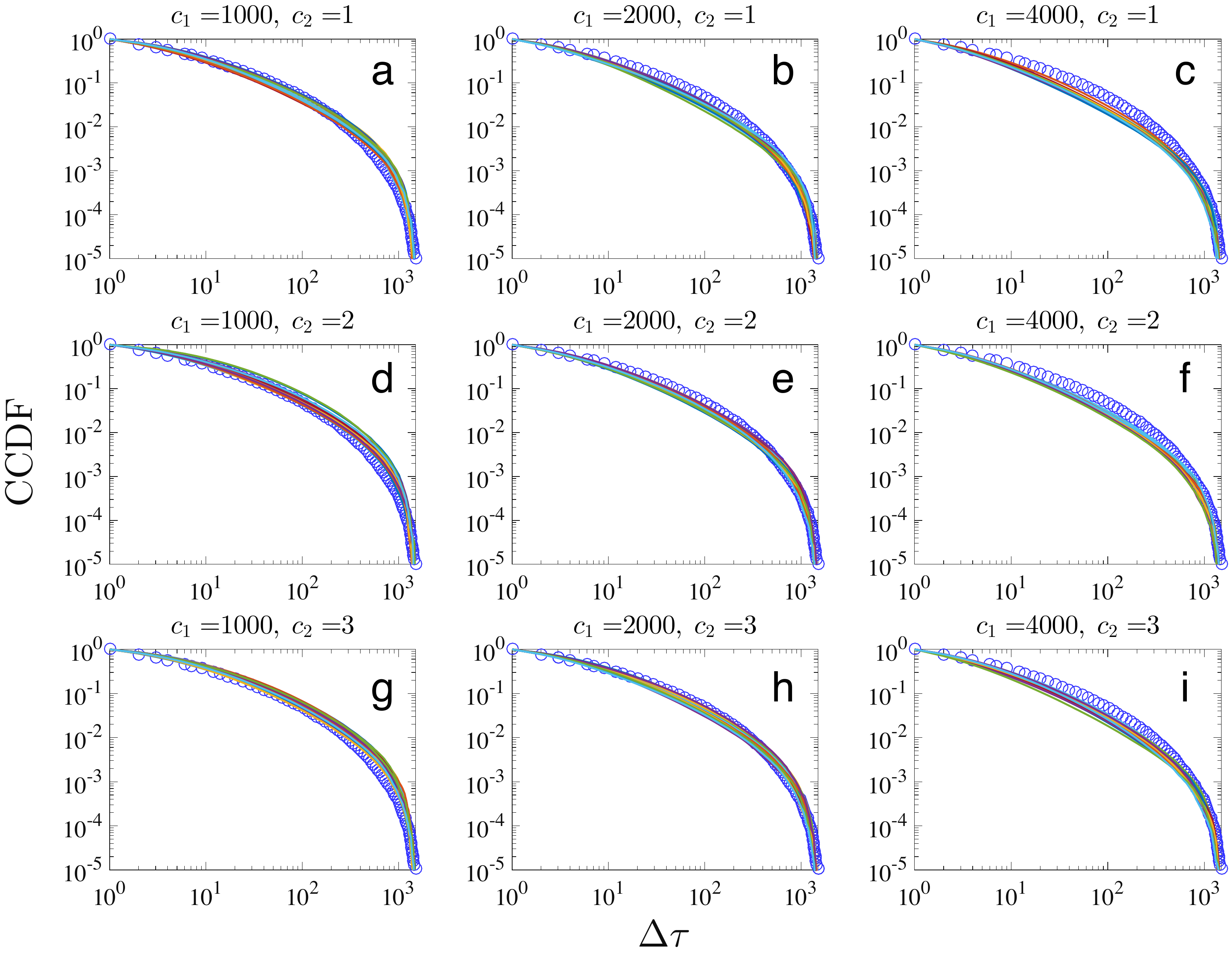}
    \end{center}
    \caption{Effect of altering the specification of reset probability $h(a)$ on the interval distribution for bank pairs. See Fig.~\ref{fig:duration_diff_resetprob} for details.}\label{fig:interval_diff_resetprob}%[do not remove: created by Nbar_durint_diff_resetprob.m]
\end{figure*}

 \clearpage

\begin{figure*}
\begin{center}
     \includegraphics[width=1.35\columnwidth]{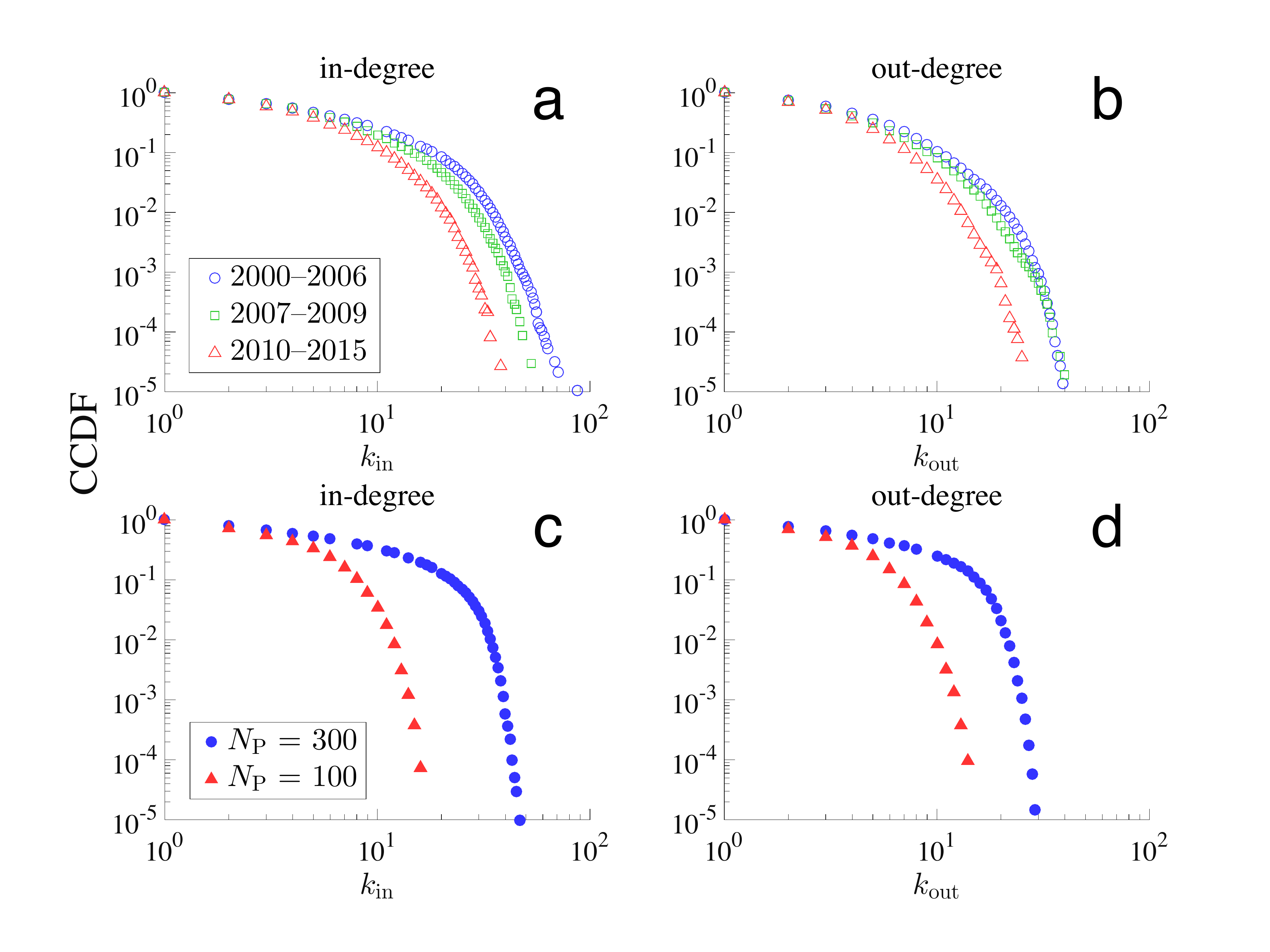}
    \end{center}
    \caption{CCDF of in- and out-degree distribution.  ({\textbf a})--({\textbf b}) The data. ({\textbf c})--({\textbf d}) The model.}\label{fig:degreedist}
\end{figure*}%[do not remove: created by Nbar_Plot_durint.m]

%%[do not remove: created by TradeProperty.m]

%[do not remove: created by Nbar_Plot_durint.m]

 \begin{figure*}
\begin{center}
      \includegraphics[width=1.5\columnwidth]{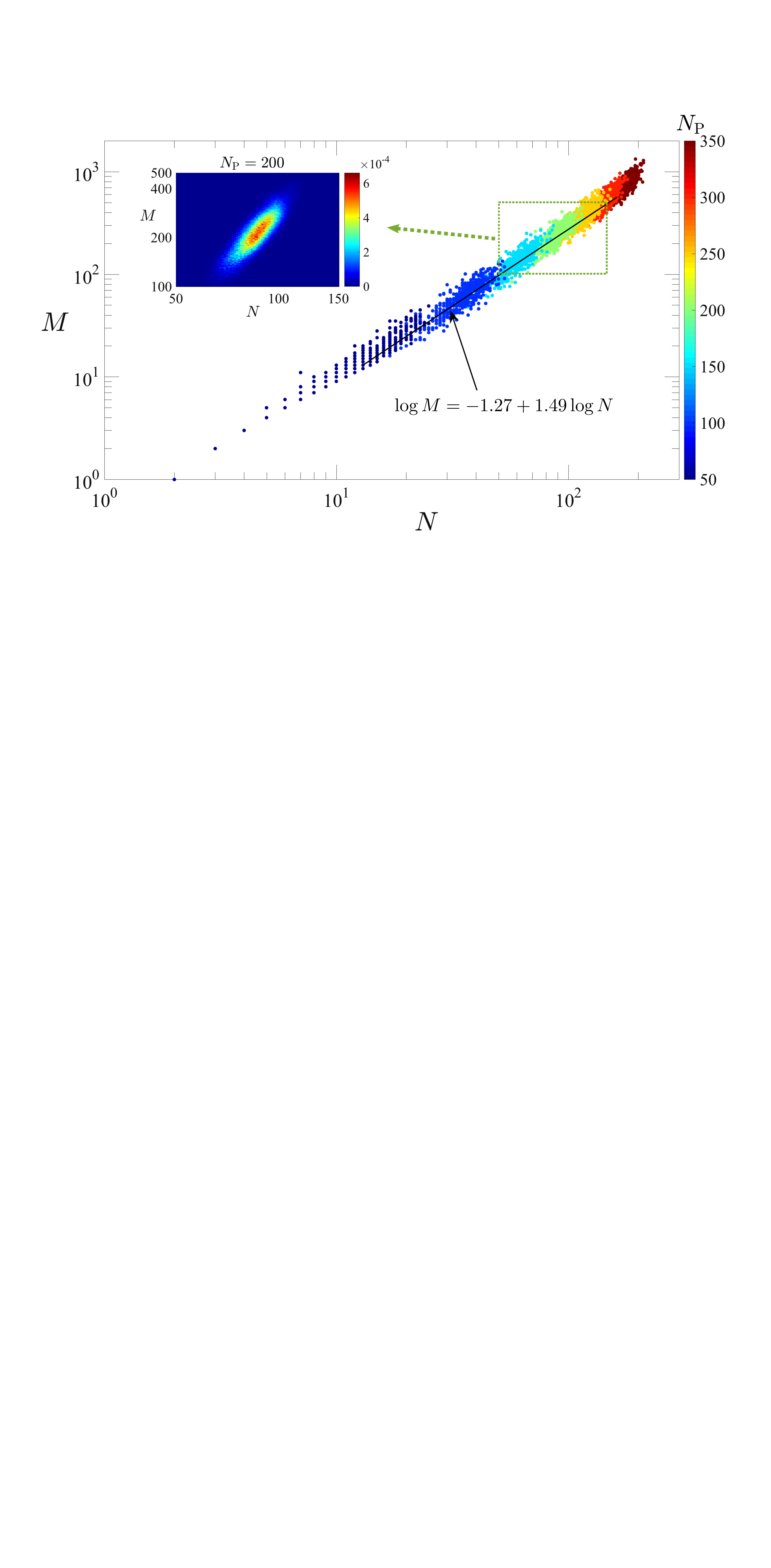} 
    \end{center}
    \caption{ Scatter plot of $(N, M)$ of networks generated with various $\Np$ values (color bar). Model networks are generated 500 times for a given $\Np$. Solid line is the empirical regression line identical to that shown in Fig.~\ref{fig:durint_scaling_model}a. \textit{Inset}: Joint conditional probability function $f(N, M | \Np)$ (color bar) with $\Np=200$.}
\label{fig:MLfunc_scaling}
\end{figure*}   

 \begin{figure*}
\begin{center}
        \includegraphics[width=1.5\columnwidth]{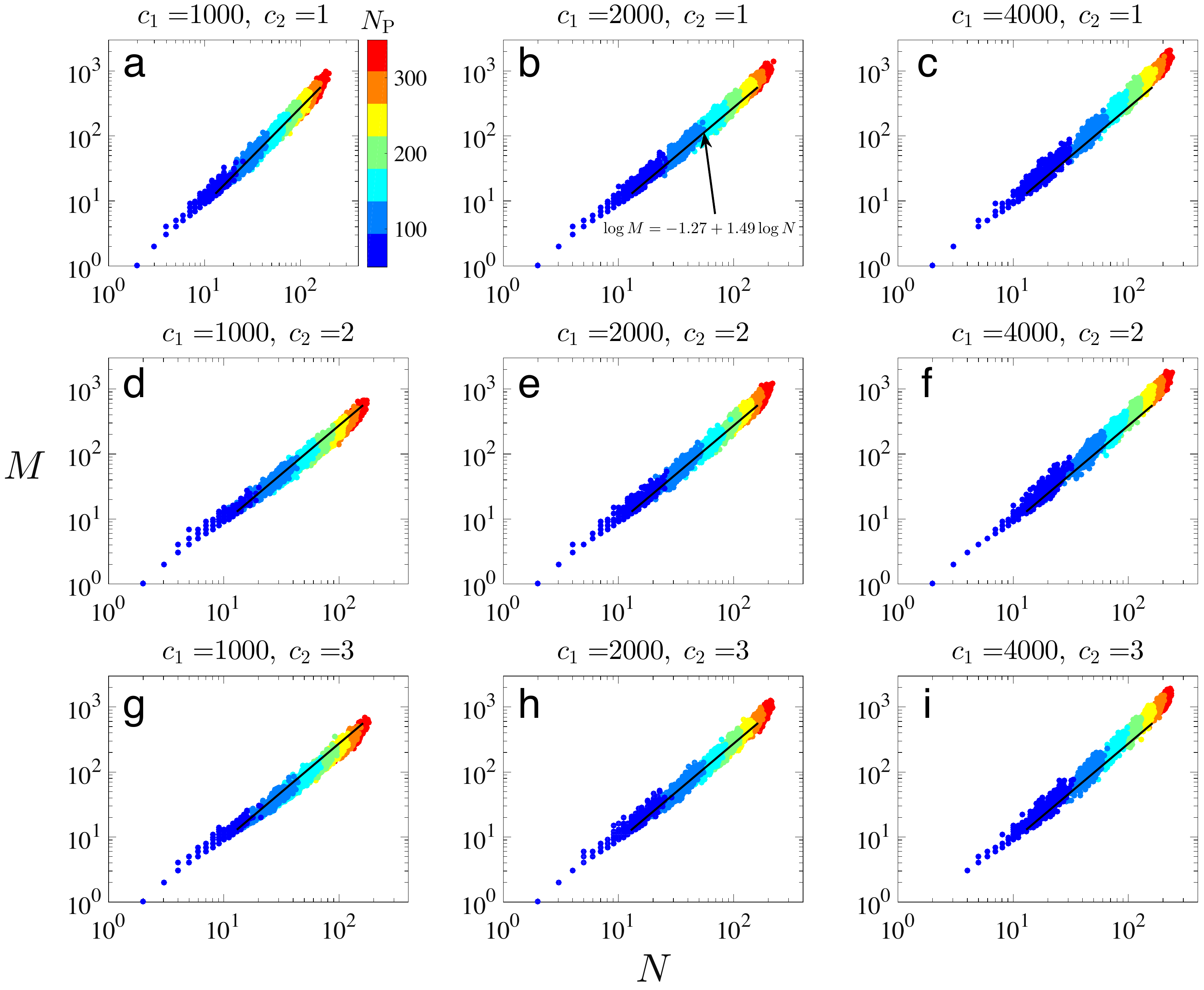}
    \end{center}
    \caption{Effect of altering the specification of reset probability $h(a)$ on scaling relation between $N$ and $M$. Parameters $c_1$ and $c_2$ are defined by $h(a) = c_{1}^{-1}a^{c_2}$. Model networks are generated 500 times for a given $\Np$.}\label{fig:scaling_diff_resetprob}%[do not remove: created by Nbar_scaling_diff_resetprob.m]
 \end{figure*}

 \begin{figure*}
 \begin{center}
     \includegraphics[width=1.5\columnwidth]{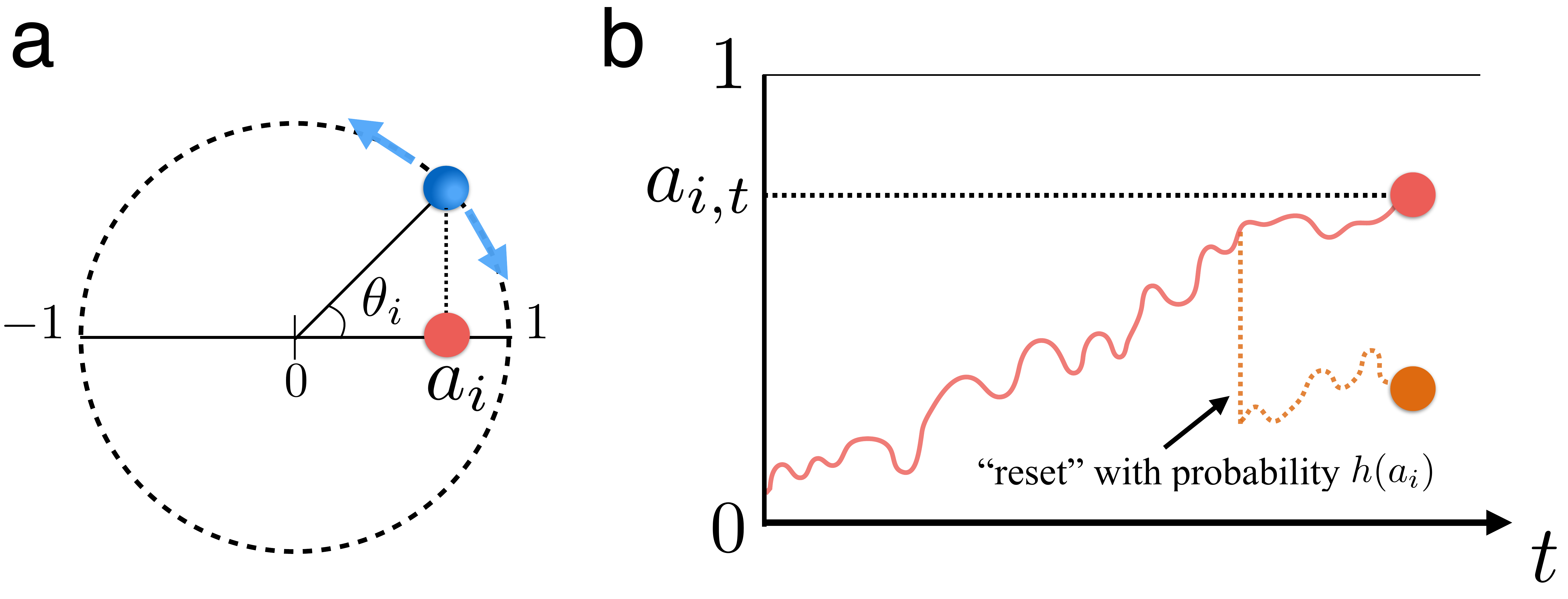}
 \end{center}
     \caption{Schematic of the activity update in the dynamical fitness model. ({\textbf a}) Circular random walk of angle $\theta_i$ on the unit circle. Activity level of bank $i$ is given by $a_i = |\cos{\theta_i}|$.  ({\textbf b}) Activity level $a_{i}$ evolves according to Eqs.~(\ref{eq:a}) and (\ref{eq:theta}) with reset probability $h(a_i)$.}\label{fig:schematic_circle}
 \end{figure*}

   \clearpage

%%%%%%%%%%%%%%%%%%Table %%%%%%%%%%%%%%%%

 \begin{table*}[t]
 \centering
            \caption{Fraction of each bank type in the data}
    \begin{tabular}{lcccc}
    \hline
    \hline
       & All & \hspace{.1cm}2000--2006 & \hspace{.1cm}2007--2009 & \hspace{.1cm}2010--2015 \\ 
       \hline
        Pure lender & 0.556 &0.553  &0.571& 0.558  \\
        Pure borrower  & 0.335 & 0.300 &0.318 & 0.380 \\
        Others &0.110 &0.147 &0.111 & 0.062  \\
        \hline
       \multicolumn{5}{l}{\footnotesize{``Pure lender" (``pure borrower") denotes the banks that lend to (borrow from) }}\\
         \multicolumn{5}{l}{\footnotesize{ but never borrow from (lend to) other banks.}}
 \end{tabular}
    \label{tab:frac_banktype}
\end{table*}

\end{document}